\documentclass[10pt]{article}

\usepackage[small]{caption}
\usepackage{graphicx}

\usepackage{amsmath,amsfonts,amssymb}
\usepackage{amsthm}
\usepackage{color}
\usepackage{authblk}

\usepackage[letterpaper,textheight=9in,textwidth=7in]{geometry}

\usepackage[skip=10pt,font=footnotesize]{caption}
\graphicspath{{Figures/}}

\newtheorem{Lemma}{Lemma}[section]
\newtheorem{Theorem}{Theorem}
\newtheorem{Proposition}[Lemma]{Proposition}
\newtheorem{Corollary}[Lemma]{Corollary}
\newtheorem{Remark}[Lemma]{Remark}
\newtheorem{Definition}[Lemma]{Definition}

\newenvironment{Proof}%
 {\begin{trivlist} \item[]{\bf Proof. }}%
 {\hspace*{\fill}$\rule{.4\baselineskip}{.4\baselineskip}$\end{trivlist}}

 {\begin{trivlist}\item[]\textbf{Acknowledgments.}}{\end{trivlist}}

\makeatletter\@addtoreset{figure}{section}\makeatother

\makeatletter \@addtoreset{equation}{section} \makeatother

\newcommand{\R}{\mathbb{R}}
\newcommand{\C}{\mathbb{C}}

\newcommand{\Z}{\mathbb{Z}}
\newcommand{\T}{\mathbb{T}}

\newcommand{\mc}[1]{\mathcal{#1}}
\newcommand{\tlR}{\tilde{\mc{R}}}
\newcommand{\tlN}{\tilde{\mc{N}}}
\newcommand{\mb}[1]{\mathbb{#1}}

\newcommand{\tl}[1]{\tilde{#1}}
\newcommand{\lp}{\left}
\newcommand{\rp}{\right}
\newcommand{\la}{\lp\langle}
\newcommand{\ra}{\rp\rangle}
\newcommand{\ba}{\begin{align}}
\newcommand{\ea}{\end{align}}
\newcommand{\fr}[2]{\frac{#1}{#2}}
\newcommand{\p}{\partial}
\newcommand{\ri}{\mathrm{i}}
 \newcommand{\rrl}{\mathrm{r/l}}
\newcommand{\rmi}{\mathrm{i}}

\newcommand{\re}{\mathrm{e}}
\newcommand{\rme}{\mathrm{e}}

\newcommand{\rmO}{\mathcal{O}}

\renewcommand{\Im}{\mathrm{Im}}

\newcommand{\rtr}{\mathrm{tr}}

\newcommand{\rs}{\mathrm{s}}

\newcommand{\rss}{\mathrm{ss}}
\newcommand{\rcu}{\mathrm{cu}}
\newcommand{\ru}{\mathrm{u}}

\newcommand{\bk}{{\mathbf k}}

\title{Pattern-forming fronts in a Swift-Hohenberg equation with directional quenching --- parallel and oblique stripes}

\author{Ryan Goh and Arnd Scheel
}

\newcommand{\Addresses}{{
  \bigskip
  \footnotesize

  R.~Goh (Corresponding author), \textsc{Department of Mathematics and Statistics, Boston University, 111 Cummington Mall, Boston,  MA 02215}\par\nopagebreak
  \textit{E-mail address}, R.~Goh: \texttt{rgoh@bu.edu}

  \medskip

  A.~Scheel, \textsc{School of Mathematics, University of Minnesota, 206 Church St. SE, Minneapolis, MN 55455}\par\nopagebreak
  \textit{E-mail address}, A.~Scheel: \texttt{scheel@umn.edu}
}}

\begin{document}

\maketitle

\abstract{
\setlength{\parindent}{0pt}
We study the effect of domain growth on the orientation of striped phases in a Swift-Hohenberg equation. Domain growth is encoded in a step-like  parameter dependence that allows stripe formation in a half plane, and suppresses patterns in the complement, while the boundary of the pattern-forming region is propagating with fixed normal velocity. We construct front solutions that leave behind stripes in the pattern-forming region that are parallel to or at a small oblique angle to the boundary.

Technically, the construction of stripe formation parallel to the boundary relies on ill-posed, infinite-dimensional spatial dynamics. Stripes forming at a small oblique angle are constructed using a functional-analytic, perturbative approach. Here, the main difficulties are the presence of continuous spectrum and the fact that small oblique angles appear as a singular perturbation in a traveling-wave problem. We resolve the former difficulty using a farfield-core decomposition and Fredholm theory in weighted spaces. The singular perturbation problem is resolved using preconditioners and boot-strapping. 
}

\setlength{\parskip}{4pt}
\setlength{\parindent}{0pt}

\section{Introduction}\label{s:intro}

We are interested in the growth of crystalline phases in macro- or mesoscopic systems, subject to directional quenching. More precisely, we are interested in systems that exhibit stable or metastable ordered states, such as stripes, or spots arranged in hexagonal lattices.  Examples of such systems arise for example in di-block copolymers \cite{PhysRevE.92.042602}, phase-field models \cite{robbins2012modeling,archer2012solidification}, and other phase separative systems \cite{foard2012survey,thomas2013probability,kopf2014emergence}, as well as in phyllotaxis \cite{pennybacker2013phyllotaxis}, and reaction diffusion systems \cite{miguez2006effect,crampin1999reaction}. Throughout, we will focus on a paradigmatic model, the Swift-Hohenberg equation
\begin{equation}\label{e:sh}
u_t = -(1+\Delta)^2 u + \mu u - u^3,
\end{equation}
where $u=u(t,x,y)\in\R$, $(x,y)\in\R^2$, $t\in\R$, subscripts denote partial derivatives, and $\Delta u = u_{xx}+u_{yy}$. It is well known that for  $\mu>0$, \eqref{e:sh} possesses  stable striped patterns $u_\mathrm{p}(k_x x + k_y y;k)$, with $u_\mathrm{p}(\theta;k) = u_\mathrm{p}(\theta+2\pi;k)$ and wave vector $\bk = (k_x,k_y)$, $k=|\bk|$ for $\mu>0$. The wave vector can be thought of as the lattice parameter of the crystalline phase, encoding its strain and orientation. 

The particular scenario of interest here is when such a system is quenched into a pattern-forming state in a growing half plane $\{(x,y)\,|\, x-ct<0\}$, choosing for instance $\mu = -\mu_0\, \mathrm{sign}\, (x-ct)$.  The main question is then if the growth process will \emph{select} an orientation or strain in the crystalline phase. Roughly speaking, our analysis establishes such a selection mechanism, for the strain, as a function of an arbitrary orientation, at least for small oblique angles. 

In a moving coordinate frame $\xi = x- ct$, \eqref{e:sh} then reads
\begin{equation}\label{e:sh0}
u_t = -(1+\Delta)^2u - \mu_0 \mathrm{sign}\,(\xi) u-u^3 + cu_\xi. 
\end{equation}
The growth process and the selection of stripes is encoded in the existence and stability of coherent structures, that is, traveling waves or time-periodic solutions to \eqref{e:sh0}. 

Stripes parallel to the boundary in a comoving frame are of the form $u_\mathrm{p}(k_x x;k_x)=u_\mathrm{p}(k_x(\xi+ct);k_x)$, hence time-periodic. 
The simplest solutions enabling the creation of such stripes are therefore of the form $u(\zeta,\tau)$, $\tau=\omega t$, $\zeta=k_x\xi$, $\omega=ck_x$, solving the boundary-value problem on $(\zeta,\tau)\in\R^2$,
\begin{equation}\label{e:par}
\left\{\begin{array}{rl}-\omega u_{\tau} -(k_x^2 \partial_\zeta^2 + 1)^2u+\mu(\zeta)u-u^3 + \omega u_\zeta&=0\\
 u(\zeta,\tau)-u(\zeta,\tau+2\pi)&=0\\
\lim_{\zeta\rightarrow\infty} u(\zeta,\tau)&=0\\
\lim_{\zeta\rightarrow-\infty} \left(u(\zeta,\tau)-u_\mathrm{p}(\tau+\zeta;k_x)\right)&=0,
\end{array}\right.
\end{equation}
for some $k_x$. 

Obliques stripes $u_\mathrm{p}(k_x x + k_y y;k)=u_\mathrm{p}(k_x \xi +k_y y + c k_x t;k)$ are stationary in a vertically comoving frame $\tau=-(k_y y + ck_x t)$. The simplest solutions enabling the creation of oblique stripes are therefore of the form $u(\zeta,\tau)$, with $\zeta=k_x\xi$, $\tau=-(k_y y + \omega t)$,  $\omega=ck_x$, and solve
\begin{equation}\label{e:obl}
\left\{\begin{array}{rl}
-\omega u_{\tau}  -(k_x^2 \p_\zeta^2 + k_y^2 \p_\tau^2 + 1)^2 u + \mu(\zeta) u - u^3 + \omega u_\zeta&=0\\
u(\zeta,\tau)-u(\zeta,\tau+2\pi)&=0\\
\lim_{\zeta\rightarrow\infty}\,\, u(\zeta,\tau)&=0 \\
\lim_{\zeta\rightarrow-\infty} \left(u(\zeta,\tau)-u_\mathrm{p}(\tau+\zeta;k)\right)&=0,
\end{array}\right.
\end{equation}
for some $k_x$. Note that, formally letting $k_y\to 0$, the problem \eqref{e:obl} limits on \eqref{e:par}. The difficulty in this limiting process is two-fold. First, the perturbation is singular in that the highest derivatives  in $\tau$ vanish at $k_y=0$. Second, the linearization  $\mathcal{L}$ of \eqref{e:par} at a solution $u_\mathrm{tr}^*$, 
\begin{equation}\label{e:lin}
\mathbb{L}v:=
-\omega v_{\tau} -(k_x^2 \partial_\zeta^2 + 1)^2v+\mu(\zeta)v-3(u_\mathrm{tr}^*)^2 v+ \omega v_\zeta,
\end{equation}
is not Fredholm as a closed and densely defined operator on  $L^2(\R\times \mb{T})$, say, where $\mb{T}= \R/2\pi\Z$. This can be readily seen noticing that $\partial_\tau u_\mathrm{tr}^*$ belongs to the kernel but does not converge to zero at infinity, such that a simple Weyl sequence construction shows that the range is not closed. 

It turns out that Fredholm properties can be recovered by choosing exponentially weighted function spaces. We therefore introduce the space 
\begin{equation}\label{e:exp}
L^2_\eta(\R\times \mb{T})=\left\{u(\zeta,\tau)\in L^2_\mathrm{loc}(\R\times\mb{T})\ \mid \ \rme^{\eta|\zeta|}u\in L^2(\R\times\mb{T})\right\},
\end{equation} 
and consider $\mathbb{L}$ as a closed operator on this exponentially weighted space with small weights $\eta\sim 0$. 

\begin{Definition}[Non-degenerate parallel stripe formation]\label{d:p}
We say that a solution $u_\mathrm{tr}^*$ of \eqref{e:par} is non-degenerate if the linearization $\mathbb{L}$ is Fredholm of index 0 in the weighted space $L^2_\eta$ for all $\eta<0$, sufficiently small, and $\lambda=0$ is algebraically simple as an eigenvalue in these spaces. Moreover, the asymptotic periodic patterns are stable with respect to coperiodic perturbations, with a simple zero eigenvalue of the coperiodic linearization induced by translations. 
\end{Definition}
We refer the reader to \cite{sandstede2004defects} for background and a spatial dynamics illustration motivating such non-degeneracy conditions. Note also that this choice of exponential weights allows for exponential growth of functions and is hence generally ill-suited for nonlinear analysis.

\paragraph{Main results.} We are now in a position to state our results. 
Our first result is concerned with a
singular perturbation in the presence of essential spectrum. 

\begin{Theorem}[Parallel $\Longrightarrow$ oblique stripe formation]\label{t:2}
Suppose there exists a solution $(u_\mathrm{tr}^*,k_x^*)$ of \eqref{e:par} for some fixed $c>0$, $\mu_0>0$, forming parallel stripes. Suppose furthermore that the solution is non-degenerate as stated in Definition \ref{d:p}.  Then there exists a family of solutions $(u_\mathrm{tr},k_x)$ to \eqref{e:obl}, depending on $k_y\sim 0$, sufficiently small, forming oblique stripes. At $k_y=0$, this family coincides with $(u_\mathrm{tr}^*,k_x^*)$.  The dependence of $k_x$ and of the solutions $u_\mathrm{tr}$ on $k_y$, measured in $C^0_\mathrm{loc}(\R\times\mb{T})$, is of class $C^2$. At leading order, the wavenumber satisfies the expansion $k_x(k_y) = k_x^* - \fr{b_y}{c}k_y^2 + \mc{O}(k_y^4)$ with $b_y$  defined in \eqref{e:coef}, below.
\end{Theorem}

Our second result shows that the assumptions of Theorem \ref{t:2} hold for $\mu_0>0$, sufficiently small. 

\begin{Theorem}[Existence of parallel stripe formation]\label{t:1}
For all $\mu_0>0$ sufficient small and $0<c<c_*(\mu_0)$, $c_*(\mu_0)= 4\sqrt{\mu_0}+\rmO(\mu_0)$, there exists a $k_x(c)$ and solution $u_\mathrm{tr}^*$ to \eqref{e:par} that is non-degenerate in the sense of Definition \ref{d:p}.
\end{Theorem}

Together, these two results establish the existence of crystallization fronts forming stripes with small oblique angle to the interface $x=ct$. In particular, crystallization fronts select wavenumbers transverse to the interface, depending on prescribed wavenumbers parallel to the interface. Using that $k\sim 1$, one can equivalently parameterize strain in the crystalline phase as a function of grain orientation, that is, growth selects strain but not orientation in this case of near-parallel orientation; see Remark \ref{r:kexpn}.

We expect non-degeneracy as in Definition \ref{d:p} to hold generically for parallel stripe formation. In particular, we expect Theorem \ref{t:2} to apply to parallel stripe formation at finite amplitude, $\mu_0$ not necessarily small, or in other systems exhibiting striped phases, as described above. Theorem \ref{t:1}, on the other hand, is intrinsically focused on small amplitudes. It seems difficult to obtain existence results of this type at finite amplitude. We expect, however, that the methods from Theorem \ref{t:1} could be adapted to find solutions to \eqref{e:obl}, at small amplitude. We chose the alternative approach from Theorem \ref{t:2} in order to illustrate the robust continuation from parallel to oblique stripes, independent of small amplitude assumptions and, to some extent, specific model problems.

\paragraph{Techniques.}
We next comment on technical aspects of the proofs of Theorems \ref{t:2} and \ref{t:1}.

The proof of Theorem \ref{t:1} pairs the somewhat classical techniques of center manifold reduction and normal forms (see \cite{eckmann1991propagating})  with heteroclinic matching techniques such as invariant foliations and Melnikov theory in an infinite dimensional setting. In particular, we perform a center manifold reduction in the spatial dynamical systems for  $\xi<0$ and $\xi>0$, separately.  The parallel striped front is then constructed as a heteroclinic orbit from the intersection of the unstable manifold of a periodic orbit in the $\xi<0$-center manifold with the stable manifold of the origin in the $\xi>0$-center manifold. Since locally the two center manifolds only intersect at the origin, we construct the heteroclinic by finding intersections of the center-unstable manifold of the periodic orbit in the $\xi<0$ dynamics with the stable manifold of the origin in the $\xi>0$ dynamics. We use invariant foliations to reduce the infinite-dimensional nature of the problem, allowing us to project the dynamics onto one of the center manifolds and obtain a leading order intersection. We then use Melnikov theory and transversality arguments to obtain the desired heteroclinic. Transversality of the intersection implies non-degeneracy as stated in Definition \ref{d:p}.

Traditionally, the two main difficulties in proving a result like Theorem \ref{t:2} have been addressed using spatial dynamics.  To address the neutral continuous spectrum induced by the asymptotic roll state, one typically uses spatial dynamics in the $\zeta$-direction.  That is one formulates the equation as an (ill-posed) dynamical system with evolutionary variable $\zeta$ and studies roll states as periodic orbits and fronts as heteroclinic orbits. Essential spectrum corresponds to the lack of hyperbolicity of periodic orbits, and is resolved by focusing on strong stable foliations.

To address the singular limit $k_y=0$, one might use spatial dynamics in the direction $\tau$ along the growth interface, formulating the equation as a fast-slow dynamical system in the $\tau$-direction \cite{sandstede2004defects,risler}. One would then try to study bifurcations using a center-manifold reduction or the geometric methods pioneered by Fenichel  \cite{fenichel1979geometric}.

Combining these two difficulties seems beyond the scope of spatial dynamics techniques, and we therefore resort to a more direct functional-analytic approach. To overcome the singularly perturbed nature of the problem, we use an approach similar to \cite{rademacher2007saddle}, preconditioning the problem with a constant-coefficient linear operator before applying the Implicit Function Theorem. To resolve the difficulties caused by the continuous spectrum, we use an ansatz of the form
\begin{equation}\label{e:chi}
u = w(\zeta,\tau) + u_\mathrm{tr}^*(\zeta,\tau) + \chi(\zeta) \lp[ u_\mathrm{p}(\zeta+\tau;k) - u_\mathrm{p}(k_x^*\zeta/k_x+\tau;k_x^*)  \rp], 
\end{equation}
to decompose the far-field patterns at $\zeta=-\infty$ from the ``core" patterns near the interface at $\zeta=0$. Here, $\chi$ is a smooth, monotone function with $\chi\equiv1$ for $\xi<-2$ and $\chi\equiv 0$ for $\xi>-1$. Having accounted in this way for changes in the farfield wavenumber, we may require the correction $w$ to be exponentially localized. We then solve in exponentially localized spaces, where the linearization turns out to be Fredholm of index -1, using the free parameter  $k_x$ as a variable to account for the cokernel. As a result, we obtain $w$ and $k_x$ as functions of the remaining free parameter $k_y$ using the Implicit Function Theorem after careful preconditioning; see Section \ref{s:po}.  Regularity in $k_y$ is  obtained via a bootstrapping procedure; see Section \ref{ss:boot}.

\paragraph{Outline.}

We prove Theorems \ref{t:2} and \ref{t:1} in Section \ref{s:po} and \ref{s:on}, respectively. We conclude with a discussion of applications, extensions, and future directions in Section \ref{s:dis}.

 \paragraph{Acknowledgements.} Research partially supported by the National Science Foundation through the grants NSF-DMS-1603416 (RG), and NSF-DMS-1612441, NSF-DMS-1311740 (AS), as well as a UMN Doctoral Dissertation Fellowship (RG).  RG would like to thank C. E. Wayne for useful discussions about this work, as well as the Institute for Mathematics and its Applications for its kind hospitality during a weeklong visit where some of this research was performed.

\section{From parallel to oblique stripes --- proof of Theorem \ref{t:2}}
\label{s:po}
We prove Theorem \ref{t:2}. Section \ref{s:prep} collects and reinterprets information on the primary profile $u^*_\rtr$ and the linearization. We then  describe the functional-analytic setup, in particular the farfield-core decomposition, Section \ref{ss:fcnl}. Section \ref{s:reg} introduces the second key ingredient to the proof, a nonlinear preconditioning to set up the Implicit Function Theorem. Section \ref{ss:pfex} concludes the existence proof and Section \ref{ss:boot} establishes differentiability in $k_y$. 

\subsection{Properties of the parallel trigger and its linearization}\label{s:prep}

We establish smoothness, exponential convergence, and some properties of the linearization. First, notice that $u^*_\rtr$ solves a pseudo-elliptic equation, such that $\partial_\tau u$ and $\partial_\zeta^4u$ belong to $L^\infty$, with a jump at $\zeta=0$. For $\zeta\neq 0$, $u^*_\rtr$ can readily seen to be smooth. By assumption, $u^*_\rtr$  converges to the periodic pattern $u_\mathrm{p}(\zeta+\tau;k^*)$ as $\zeta\rightarrow-\infty$. 

By translation invariance in $\tau$, $\partial_\tau u^*_\rtr$ is bounded and belongs to the kernel of the linearized equation in $L^2_\eta$, for all $\eta<0$, but not for $\eta\geq0$, since $k_x\neq 0$. 

\begin{Lemma}[Fredholm crossing]\label{l:fc}
The operator $\mathcal{L}$ from \eqref{e:lin} is Fredholm of index -1 in $L^2_\eta(\R\times\T)$ for $\eta>0$, sufficiently small, with trivial kernel. 
\end{Lemma}
\begin{Proof}
We rely on the characterization of Fredholm indices using Fredholm borders; see \cite{sandstede2001structure,sandstede2008relative,fiedlerscheel}. Since the asymptotic state at $\xi=+\infty$ is linearly stable, its Morse index at $\xi=-\infty$ can be calculated through a homotopy as follows. We linearize at the asymptotic periodic pattern and find, including a spectral homotopy parameter $\lambda$, 
\[
-\omega (u_\tau-u_\zeta)-(k_x^2\partial_\zeta^2+1)^2u+\mu_0 u - 3 u_\mathrm{p}^2(\zeta+\tau)u=\lambda u.
\]
A Floquet-Bloch ansatz $u(\zeta,\tau)=\rme^{\rmi\ell\tau}\rme^{\nu_x\zeta}w(\zeta+\tau)$, with $w(z)=w(z+2\pi)$, yields the periodic boundary value problem 
\[
\lambda u = -\omega (\rmi \ell -\nu_x)-(k_x^2(\partial_z+\nu_x)^2+1)u + \mu_0 u -3u_\mathrm{p}^2(z)u =:\mathbb{L}_\mathrm{p}(\nu_x)u-\omega(\rmi\ell - \nu_x) u,\quad z\in(0,2\pi).
\]
We are interested in spatial eigenvalues $\nu_x\in\C$, crossing the imaginary axis, that is, $\nu_x\in\rmi\R$. In this case, $\mathbb{L}_\mathrm{p}(\nu_x)$ is Hermitian, such that for homotopies $\lambda\geq 0$ we necessarily find $\nu_x=\rmi\ell$ and $\mathbb{L}_\mathrm{p}(\rmi\ell)w=\lambda w$. Restricting to the fundamental Floquet domain $\Im\nu_x\in[0,1)$, we further conclude $\nu_x=\ell=0$. This however is impossible for $\lambda>0$ by the assumption of coperiodic stability, and it implies that $w=u_\mathrm{p}'$ up to scalar multiples for $\lambda=0$, $\nu_x=\ell=0$. Inspecting multiplicity of this spatial Floquet multiplier $\nu_x=0$, one is looking for a generalized eigenfunction $w$ solving $\mathbb{L}_\mathrm{p}'(0)u_\mathrm{p}'+\omega u_\mathrm{p}'+\mathbb{L}_\mathrm{p}(0)w=0$ for $w$, where $\mathbb{L}_\mathrm{p}'$ denotes the derivative with respect to $\nu_x$. By the quadratic  dependency on $\nu_x$, this reduces to $\mathbb{L}_\mathrm{p}(0)w=\omega u_\mathrm{p}'$, which in turn is impossible since $\mathbb{L}_\mathrm{p}(0)$ is self-adjoint. This proves that there is precisely one zero Floquet exponent for $\lambda=0$ and no Floquet exponent crossings for $\lambda>0$, which implies that $\mathcal{L}$ is Fredholm of index 0 for $\eta<0$, small, and Fredholm of index -1 for $\eta>0$, small.  Since the eigenfunction at $\lambda=0$ is not decaying as $\xi\to -\infty$, the kernel is trivial for $\eta>0$. 
\end{Proof}

\begin{Lemma}\label{l:exp}
There are $C,\delta>0$, such that
\begin{equation}\label{e:ecvg}
\|
\chi(\zeta+\zeta_0)
\lp( 
u^*_\rtr(\zeta/k_x^*,\tau)-u_\mathrm{p}(\zeta+\tau;k_x^*) 
\rp) 
\|_{H^k(\R\times\mb{T})}\leq C \rme^{-\delta \zeta_0}
\end{equation}
as $\zeta_0\to +\infty$, where $\chi$ is as in \eqref{e:chi}.
\end{Lemma}
\begin{Proof}
We can rely on spatial dynamics; see for isntance \cite{sandstede2004defects}. The asymptotic periodic orbit is hyperbolic up to the neutral Floquet exponent generated by translations, as seen in Lemma \ref{l:fc}. Its local stable manifold is therefore given by the union of strong stable fibers, thus implying exponential convergence. Since the spatial dynamics can be formulated in spaces of arbitrary regularity, convergence is exponential in spaces with higher derivatives, too. %

\end{Proof}

\subsection{Setup and farfield-core decomposition}\label{ss:fcnl}
We start setting up our fixed point argument by performing a functional analytic farfield-core decomposition. We start from \eqref{e:obl}, 
\begin{equation}\label{e:sh0sc}
0= -(1 +(k_x\p_\zeta)^2+(k_y\p_\tau)^2)^2 u + \mu(\zeta) u - u^3 + ck_x( u_\zeta -  u_{\tau}),\qquad (\zeta,\tau)\in \R\times\mb{T},
\end{equation}
with the appropriate boundary conditions stated there. Recall that $\chi$ is a smooth cut-off function with $\chi \equiv 1$ for all $\xi<-2$ and support contained in $(-\infty,-1]$.   Our ansatz is of the form 
\begin{equation}\label{e:ansatz}
u = w(\zeta,\tau) + u^*_\rtr\lp( \frac{k_x^*}{k_x}\zeta,\tau\rp)+ \chi(\zeta) \lp(u_\mathrm{p}\lp(\zeta+\tau;k\rp)  - u_\mathrm{p}\lp( \frac{k_x^*}{k_x}\zeta+\tau;k_x^*\rp)  \rp),
\end{equation}
where $k^2=k_x^2+k_y^2$. We consider perturbations $w\in L_\eta^2$ defined in \eqref{e:exp}, with $\eta>0$ sufficiently small.  
Note that we introduced parameter dependence into the trigger front $u_\rtr^*$ by simply scaling appropriately with $k_x$, such that $u_\rtr^*$ solves 
$\mc{L}(k_x,0) u_\rtr^* + f(u_\rtr^*) + c(k_x^* - k_x)\p_\tau u_\rtr^* = 0$ for all $k_x\neq0$.

Let us first consider the non-regularized nonlinear map.  Inserting the ansatz \eqref{e:ansatz} into \eqref{e:obl}, and setting
\begin{align}\label{e:set}
\mc{L}(k_x,k_y) &= -(1+\Delta_{k_x,k_y})^2 + ck_x(\p_\zeta-\p_\tau),\quad f(u) =\mu(\zeta) u -  u^3, \quad \Delta_{k_x,k_y}:=(k_x\p_\zeta)^2 + (k_y\p_\tau)^2, \,\,\notag\\
&\Phi(\bk)(\zeta,\tau) = u_\mathrm{p}(\zeta + \tau,k) - u_\mathrm{p}(k_x^*\zeta/k_x + \tau;k_x^*),\notag
\end{align}
we find, suppressing $\tau,\zeta$ dependence, 
\begin{align}
0& = \mc{L}(k_x,k_y)(u_\rtr^* + w + \chi \Phi(\bk)) + f( u_\rtr^* + w + \chi \Phi(\bk))\notag\\
&= \mc{L}(k_x,k_y)  u_\rtr^* + \mc{L}(k_x,k_y)(\chi\Phi(\bk) + w) + f(u_\rtr^* + w + \chi \Phi(\bk)) \notag\\
&= \mc{L}(k_x,k_y)(\chi\Phi(\bk) + w) + f(u_\rtr^* + w + \chi \Phi(\bk)) -f(u_\rtr^*) - (k_x^* - k_x)c \p_\tau u_\rtr^* + (\mc{L}(k_x,k_y) - \mc{L}(k_x,0))u_\rtr^*.  \notag
\end{align}
 From this last line, we can then define the mapping 
\begin{align}
F(w;\bk):=[\mc{L}(k_x,k_y) +  f'(u_\rtr^*)]w + \mc{R}(\bk) + \mc{N}(w;\bk),
\end{align}
with the $w$-independent residual term
\begin{equation}
\mc{R}(\bk) = \mc{L}(k_x,k_y)( \chi \Phi(\bk)) + f(u_\rtr^* + \chi\Phi(\bk)) -f(u_\rtr^*) + (k_x - k_x^*)c \p_\tau u_\rtr^* + (\mc{L}(k_x,k_y) - \mc{L}(k_x,0))u_\rtr^*,
\end{equation}
and the nonlinear term
\begin{equation}
\mc{N}(w;\bk) = f(u_\rtr^* + \chi\Phi(\bk) + w) - f(u_\rtr^*+\chi\Phi(\bk)) - f'(u_\rtr^*)w.
\end{equation}
Since $f(u) = \mu u - u^3$, we have
$$
\mc{N}(w;\bk) = 3(u_\rtr^*+\chi\Phi(\bk))^2 w - 3(u_\rtr^*)^2 w + 3(u_\rtr^*+\chi\Phi(\bk))w^2+w^3.
$$
Exponential convergence of the primary trigger $u^*_\rtr$ from Lemma \ref{l:exp} implies that 
for $\bk = \bk^*:=(k^*,0)^T$, $F$ is a locally well-defined nonlinear mapping from the anisotropic Sobolev space $X_\eta := H^1(\mb{T},L^2_\eta(\R))\cap L^2(\mb{T},H^4_\eta(\R))$ to $L^2_\eta$. Indeed, one obtains $\mc{R}(\bk^*)\in L^2_\eta$ using the exponential convergence of $u_\rtr^*$ to $u_\mathrm{p}(|\bk^*|)$ and the fact that $\mc{L}(k_x,k_y)u_\mathrm{p}(|\bk|) + f(u_\mathrm{p}(|\bk|)) = 0.$  Then, since  $X_\eta$ can readily be seen to be a Banach algebra and $u_\mathrm{p}$ is bounded, we have that $\mc{N}(w;k)\in L^2_\eta$ as well.

The existence assumption of $u_\mathrm{tr}^*$ implies that $F(0,\bk^*) = 0.$  
Also, we note that the linearization of $F$ in $w$ satisfies
\begin{align}
\p_wF\Big|_{(0,\bk^*)} &= \mb{L}:=-(1+(k_x^*\p_\zeta)^2)^2 + f'(u_\rtr^*) + ck_x^*(\p_\zeta - \p_{\tau}),\notag
\end{align}
which is closed and densely defined on $L^2_\eta$. We also record for later use that
\begin{equation}
\p_{k_x}F\Big|_{(0,\bk^*)} = [\mc{L}(k_x^*,0) + f'(u_\rtr^*)] \chi\p_k\Phi(\bk_*)- c\p_\tau u_\rtr^*.\notag
\end{equation}

It is readily observed that $F$ is not continuous in $k_y^2$ as a map from $X_\eta$ to $L^2_\eta$ as it is not well-defined for $k_y\neq 0,$ due to higher-order terms such as $\p_\tau^4$ and $\p_\tau^2(1+(k_x\p_\zeta)^2)$. We will therefore regularize the equation by preconditioning with a Fourier multiplier in the next section. We conclude this setup by collecting some properties of the linearization. 

\begin{Lemma}\label{l:fred}
For all $\eta>0$ small, let $e_*$ span $\mathrm{coker}\, \mb{L}$  with $\la e_*,\p_\tau u_{\rtr}^*\ra_{L^2_{\eta}}=1$. Algebraic simplicity of the eigenvalue 0 in $L^2_\eta$, $\eta<0$, as in Definition \ref{d:p} then implies that 
\begin{equation}\label{e:coeff}
\Big\langle e_*, \p_{k_x} F\Big|_{(0,k_x^*,0)}\Big\rangle_{L^2_\eta}  \neq0.
\end{equation}
\end{Lemma}
\begin{Proof}
We argue as in \cite[Lemma 6.3]{lloyd2016continuation}. Namely, if one assumes that the above inner product is zero, then there exists $w_0\in L^2_\eta$ and $\alpha\in\R$ non-zero such that
\[
\mb{L}w_0 =\alpha\p_{k_x}F = \alpha\lp(\mb{L}(\chi \p_{k_x} \Phi(\bk^*)) - c \p_\tau u_\mathrm{tr}^*\rp).
\]
We then have 
\[
\mb{L}(w_0 - \alpha \chi \p_{k_x} \Phi(\bk^*)) = c\p_\tau u_\mathrm{tr}^*.
\]
Since $w_0$ is exponentially localized, and $\p_{k_x}u_\mathrm{p}\in L^2_{-\eta}$, we have $w_0 - \alpha \chi\p_{k_x} \Phi(\bk^*)) \neq 0$ and is a generalized eigenvector of the kernel element $\p_\tau u_\mathrm{tr}^*$ when considered in $L^2_{-\eta}$, contradicting the algebraic simplicity.
\end{Proof}

We also define the $L^2$-adjoint of $\mb{L}$ for later use, 
\begin{align}
\mb{L}_0^\mathrm{ad}: X_0\subset L^2&\rightarrow L^2\notag\\
v&\mapsto -(1+(k_x^*\p_x)^2)^2v-ck_x^*( \p_x-\p_\tau)v  +f'(u_\rtr^*)v.
\end{align}

\subsection{Regularization of the nonlinear mapping}\label{s:reg}

To prove Theorem \ref{t:2}, we shall study zeros of the regularized nonlinear mapping
\begin{equation}\label{e:mf}
\mc{F}(w;k_x,k_y^2):=\mc{M}(k_x,k_y^2) F(w;\bk): X_\eta\rightarrow X_\eta,
\end{equation}
with the regularizing operator $\mc{M}(k_x,k_y^2) := \lp(\mc{L}(\bk) - \mathrm{id}\rp)^{-1}.$  The existence and boundedness of such an inverse can be obtained in $L^2$ by studying the associated Fourier multiplier,
$$
\widehat{\mc{M}}(k_x,k_y^2)(\ri m, \ri \ell) = \lp(-1  -(1 - k_x^2\ell^2 - k_y^2 m^2)^2 + c k_x\ri(\ell - m) \rp)^{-1} ,\quad m\in \Z,\  \ell\in \R,
$$
noticing that the real part of the denominator has real part less than -1 for all $m,\ell$. One can then extend existence to the exponentially weighted space $L^2_\eta$ by the use of conjugating isomorphisms $u\mapsto \re^{\pm\eta\la x\ra}u$.  More explicitly one obtains the inverse by first obtaining the inverse on the one-sided weighted spaces $L^2_{\pm\eta,>}$, where $L^2_{\eta,>} = \{ u: \re^{\eta \xi} u\in L^2\}$ and $\p_\xi$ acts as $\p_\xi - \eta$, and then use the fact that $L^2_\eta = L^2_{\eta,>}\cap L^2_{-\eta,>}$ with equivalent norm,
$$
|| u ||_{ L^2_{\eta,>}\cap L^2_{-\eta,>}} := ||u||_{L^2_{\eta,>} }+ ||u||_{L^2_{-\eta,>}},
$$

This inverse can also be found to have the continuity properties listed in the following proposition.  For ease of notation in the following we let $\epsilon=k_y^2$, $\kappa = k_x$, and $\kappa_*= k_x^*$. Also, since they are in fact functions of $\epsilon = k_y^2$, we re-define $\mc{R}(\kappa,\epsilon) = \mc{R}(\bk)$ and $\mc{N}(w;\kappa,\epsilon) = \mc{N}(w;\bk)$.  
\begin{Proposition}\label{p:m1}
For $\epsilon\sim 0,\kappa\sim \kappa_*\neq0$, and speed $c\neq0$, the mappings $\mc{M}(\kappa,\epsilon)$ and $\p_\kappa \mc{M}(\kappa,\epsilon)$ are well-defined, bounded, and norm-continuous on  $X_\eta\times \R^2\rightarrow X_\eta$. 
\end{Proposition}
\begin{Proof}
Since we are considering $\mc{M}$ as a mapping from $X_\eta$ to itself, it suffices to prove the result on $L^2_\eta$. Furthermore we prove the result for $\eta = 0$ as the result for $0<\eta\ll1$ can be obtained as discussed above.  Hence, we consider $\mc{M}:L^2\rightarrow L^2$.  

We start by considering continuity of $\mc{M}$. Therefore let $\epsilon\to \epsilon_0,\kappa\to \kappa_0$.  For $\epsilon_0>0$, continuity is easily established using smoothness properties of the symbol. We therefore focus on the case $\epsilon_0=0, \kappa_0 = \kappa_*$. We need to show
\[
  \sup_{m,\ell}|\widehat{\mc{M}}(\kappa,\epsilon)(\ri m,\ri\ell) -\widehat{\mc{M}}(\kappa_*,0)(\ri m,\ri\ell)|\rightarrow 0.
\]  
 We decompose 
\begin{align*}
  \sup_{m,\ell}&|\widehat{\mc{M}}(\kappa,\epsilon)(\ri m,\ri\ell) -\widehat{\mc{M}}(\kappa_*,0)(\ri m,\ri\ell)|\\
  &\leq  \sup_{m,\ell}|\widehat{\mc{M}}(\kappa,\epsilon)(\ri m,\ri\ell) -\widehat{\mc{M}}(\kappa_*,\epsilon)(\ri m,\ri\ell)|+ \sup_{m,\ell}|\widehat{\mc{M}}(\kappa_*,\epsilon)(\ri m,\ri\ell) -\widehat{\mc{M}}(\kappa_*,0)(\ri m,\ri\ell)|=:\mathrm{(I)}+\mathrm{(II)}.
\end{align*}  
To estimate $\mathrm{(I)}$, we first define 
$$
p_{\kappa,\epsilon}(m,\ell) = -1 - (1 - \kappa^2 \ell^2 - \epsilon m^2)^2 + c \kappa \ri (m+\ell),
$$
 so that 
\begin{align}
\lp|\widehat{\mc{M}}(\kappa,\epsilon)(\ri m,\ri\ell) -\widehat{\mc{M}}(\kappa_*,\epsilon)(\ri m,\ri\ell)\rp|&=
\lp|\fr{p_{\kappa_*,\epsilon}(m,\ell) - p_{\kappa,\epsilon}(m,\ell)}{p_{\kappa_*,\epsilon}(m,\ell)p_{\kappa,\epsilon}(m,\ell)}\rp|\notag\\
&= |\kappa - \kappa_*| \lp| \fr{c\ri(\ell-m) - (\kappa_*^2 + \kappa^2)(\kappa_*+\kappa) \ell^4 - 2 \epsilon m^2 \ell^2(\kappa_*+\kappa)}{p_{\kappa,\epsilon}(m,\ell)p_{\kappa_*,\epsilon}(m,\ell)}   \rp|\notag\\
&\lesssim |\kappa - \kappa_*|  \fr{\lp|c\ri(\ell-m) - 4\kappa_*^3\ell^4 - 4\kappa_* \epsilon m^2 \ell^2 \rp|}{|p_{\kappa_*,\epsilon}(m,\ell)|^2}\label{e:Mkk},
\end{align}
where in the last line we used the fact that $\kappa_*\neq0$.  Then
$$|p_{\kappa_*,\epsilon}(m,\ell)^2| = |p_{\kappa_*,\epsilon}(m,\ell)|^2 = (1+(1-\kappa^2_*\ell - \epsilon m^2)^2)^2 + c^2 \kappa_*^2(\ell - m)^2,
$$ can be bounded from below by $1+\ell^8 + \epsilon\ell^6 m^2 +\epsilon^2 m^4$ up to a constant independent of $\epsilon$.  It then readily follows that the above quotient is bounded uniformly in $(\kappa - \kappa_*)$ so that 
\begin{align}
\lp|\widehat{\mc{M}}(\kappa,\epsilon)(\ri m,\ri\ell) -\widehat{\mc{M}}(\kappa_*,\epsilon)(\ri m,\ri\ell)\rp|&\lesssim |\kappa - \kappa_*|.
\end{align}

For $\mathrm{(II)}$ we further decompose
\begin{align}
\widehat{\mc{M}}(\kappa,\epsilon)(\ri m,\ri\ell)  - \widehat{\mc{M}}(\kappa,0)(\ri m,\ri\ell)
&= \fr{p_{\kappa_*,\epsilon}(m,\ell) - p_{\kappa_*,0}(m,\ell)}{p_{\kappa_*,\epsilon}(m,\ell)p_{\kappa,\epsilon}(m,\ell)}   \notag\\
&=
\fr{2\epsilon m^2(1-\kappa^2\ell^2)}{p_{\kappa_*,\epsilon}(m,\ell)p_{\kappa,\epsilon}(m,\ell) } 
-\fr{\epsilon ^2 m^4}{p_{\kappa_*,\epsilon}(m,\ell)p_{\kappa,\epsilon}(m,\ell)} =: \mathrm{(III)} + \mathrm{(IV)}.
\end{align}

For $\mathrm{(III)}$ we scale $\tl m = \epsilon^{1/3} m$ and define $\Omega_d = \{ |\tl m|^2 + |\ell|^2 > d\}$.  By studying the real and imaginary parts of the product $p_{\kappa_*,\epsilon}(m,\ell)p_{\kappa_*,0}(m,\ell)$, one finds for $(\tl m, \ell)\in \Omega_d$ and $d$ sufficiently large 
$$
|p_{\kappa_*,\epsilon}(m,\ell)p_{\kappa_*,0}(m,\ell)| \gtrsim 1 + \ell^8 + |c\kappa \tl m|(\ell^2  \tl m^2 + \epsilon^{1/3} \tl m^4),
$$ so that 
\begin{align}
|\mathrm{(III)}|&= \lp|\fr{2\epsilon^{1/3} \tl m^2(1-\kappa^2\ell^2)}{p_{\kappa_*,\epsilon}(m,\ell)p_{\kappa,\epsilon}(m,\ell) }\rp|\notag\\
&\lesssim  \epsilon^{1/3}\fr{2 \tl m^2\kappa^2\ell^2}{ 1 + \ell^8 + |c\kappa|\ell^2  |\tl m|^3 + \epsilon^{1/3} |\tl m|^5}\notag\\
&\leq C_d\epsilon^{1/3},
\end{align}
with constant $C_d>0$, independent of $\epsilon$. Note that here we have used the fact that $\kappa_*$ and $c$ are both non-zero.

On the complement $\Omega^\mathrm{c}_d$, one bounds the denominator from below by 1 so that 
\begin{align}
|\mathrm{(III)}|&\leq |2\epsilon^{1/3}\tl m^2(1 - \kappa^2\ell^2)|\notag\\
&\lesssim \epsilon^{1/3} (|\tl m|^2 + |\ell|^2)\notag\\
&\leq C_d' \epsilon^{1/3}.
\end{align}
A similar argument is used to bound $\mathrm{(IV)}$, this time with the scaling $\tl m = \epsilon^{2/5}m$. Indeed, for $(\tl m,\ell)\in \Omega_d$ we have 
\begin{align}
\mathrm{(IV)} =& \lp| \fr{2\epsilon^{2/5} \tl m^4}{p_{\kappa_*,\epsilon}(m,\ell)p_{\kappa,\epsilon}(m,\ell) }  \rp|\notag\\
\lesssim& \epsilon^{2/5} \fr{ \tl m^4}{ 1 + \ell^8 + |c\kappa_*||\tl m|^5}.
\end{align}

Combining the bounds for $\mathrm{(III)}$ and $\mathrm{(IV)}$ we conclude that the term $\mathrm{(II)}$ converges to zero as $(\kappa,\epsilon)\rightarrow(\kappa_*,0)$ as desired.

We next turn to the derivative, $\p_\kappa \mc{M}$.  First $\p_\kappa\mc{M}$ exists in a neighborhood of $(\kappa_*,\epsilon)$ and has the expected Fourier multiplier 
$$
\widehat{\p_k \mc{M}}(\kappa,\epsilon) = - \fr{\p_\kappa p_{\kappa,\epsilon}(m,\ell)}{p_{\kappa,\epsilon}(m,\ell)^2},
$$
where $p$ is defined above. This is obtained by observing
\begin{align}
&\sup_{m,\ell} \lp|  \widehat{\mc{M}}(\kappa,\epsilon) - \widehat{\mc{M}}(\kappa_0,\epsilon) - \fr{\p_\kappa p_{\kappa_0,\epsilon}(m,\ell)}{p_{\kappa_0,\epsilon}(m,\ell)^2}(\kappa - \kappa_0)\rp|\cdot|\kappa - \kappa_0|^{-1}\notag\\
&= \sup_{m,\ell} \lp|  \fr{2(\kappa + \kappa_0)\ell^2 - (\kappa^2+\kappa^2_0)(\kappa+\kappa_0)\ell^4 - 2(\kappa_* + \kappa)\epsilon\ell^2 m^2 + c\ri(\ell-m)}{p_{\kappa,\epsilon}(m,\ell)p_{\kappa_0,\epsilon}(m,\ell)}  - \fr{4\kappa_0(1-\kappa_0^2\ell^2-\epsilon m^2)\ell^2 + c\ri (\ell - m)}{p_{\kappa_0,\epsilon}(m,\ell)^2}\rp| \rightarrow 0
\end{align}
as $\kappa \rightarrow\kappa_0,$ for any $\epsilon$ near 0 and $\kappa_0$ near $\kappa_*$. This convergence is obtained using similar estimates as in \eqref{e:Mkk}. 

To prove continuity in $(\kappa,\epsilon)$, we proceed in the same way as above, aiming to show
$$
\sup_{m,\ell} \lp|  \widehat{\p_\kappa \mc{M}}(\kappa,\epsilon)(\ri m,\ri\ell) - \widehat{\p_\kappa \mc{M}}(\kappa_*,0)(\ri m,\ri\ell)\rp| \rightarrow0,\quad\text{as}\,\, (\kappa,\epsilon) \rightarrow(\kappa_*,0),
$$
and thus once again decompose
\begin{align}
&\sup_{m,\ell}\lp|  \widehat{\p_\kappa \mc{M}}(\kappa,\epsilon)(\ri m,\ri\ell) - \widehat{\p_\kappa \mc{M}}(\kappa_*,0)(\ri m,\ri\ell)\rp|\notag\\
&\leq  \sup_{m,\ell}| \widehat{\p_\kappa \mc{M}}(\kappa,\epsilon)(\ri m,\ri\ell) - \widehat{\p_\kappa \mc{M}}(\kappa_*,\epsilon)(\ri m,\ri\ell)|+ \sup_{m,\ell}| \widehat{\p_\kappa \mc{M}}(\kappa_*,\epsilon)(\ri m,\ri\ell) - \widehat{\p_\kappa \mc{M}}(\kappa_*,0)(\ri m,\ri\ell)|=:\mathrm{(I)}+\mathrm{(II)}.
\end{align}

It is readily found that 
\begin{align}
| \widehat{\p_\kappa \mc{M}}&(\kappa,\epsilon)(\ri m,\ri\ell) - \widehat{\p_\kappa \mc{M}}(\kappa_*,\epsilon)(\ri m,\ri\ell)| = \lp| \fr{\p_\kappa p_{\kappa,\epsilon}(m,\ell)p_{\kappa_*,\epsilon}(m,\ell)^2 - \p_\kappa p_{\kappa_*,\epsilon}(m,\ell)p_{\kappa,\epsilon}(m,\ell)^2}{p_{\kappa,\epsilon}(m,\ell)^2p_{\kappa_*,\epsilon}(m,\ell)^2}  \rp|\notag\\
&\lesssim\lp| \fr{\p_\kappa p_{\kappa,\epsilon}(m,\ell)(p_{\kappa_*,\epsilon}(m,\ell) - p_{\kappa,\epsilon}(m,\ell))(p_{\kappa_*,\epsilon}(m,\ell) + p_{\kappa,\epsilon}(m,\ell))}{p_{\kappa,\epsilon}(m,\ell)^2p_{\kappa_*,\epsilon}(m,\ell)^2}\rp| + 
\fr{\lp|\p_\kappa p_{\kappa_*,\epsilon}(m,\ell) - \p_\kappa p_{\kappa,\epsilon}(m,\ell)\rp|}{|p_{\kappa_*,\epsilon}(m,\ell)^2|}\notag\\
&\lesssim \fr{|\p_\kappa p_{\kappa,\epsilon}(m,\ell)||p_{\kappa_*,\epsilon}(m,\ell) - p_{\kappa,\epsilon}(m,\ell)|}{|p_{\kappa,\epsilon}(m,\ell)||p_{\kappa,\epsilon}(m,\ell)p_{\kappa_*,\epsilon}(m,\ell)|} +
\lp|\fr{\p_\kappa p_{\kappa_*,\epsilon}(m,\ell) - \p_\kappa p_{\kappa,\epsilon}(m,\ell)}{p_{\kappa_*,\epsilon}(m,\ell)^2}\rp|\notag\\
&= \fr{\lp|4\kappa\ell^2(1-\kappa^2\ell^2 -\epsilon m^2)+ c\ri(\ell-m)\rp| }{|p_{\kappa_*,\epsilon}(m,\ell)|}|\widehat{\mc{M}}(\kappa,\epsilon) - \widehat{\mc{M}}(\kappa_*,\epsilon)| \notag\\
&\qquad\quad+|\kappa - \kappa_*|\fr{|4\ell^2 + 4(\kappa^2 + \kappa\kappa_* + \kappa_*^2)\ell^4 + 4\epsilon m^2 \ell^2|}{|p_{\kappa_*,\epsilon}(m,\ell)|^2}\notag\\
&\lesssim |\kappa - \kappa_*|.
\end{align}

For $\mathrm{(II)}$ we proceed in a similar manner finding
\begin{align}
| \widehat{\p_\kappa \mc{M}}&(\kappa_*,\epsilon)(\ri m,\ri\ell) - \widehat{\p_\kappa \mc{M}}(\kappa_*,0)(\ri m,\ri\ell)|= \lp|\fr{\p_\kappa p_{\kappa_*,\epsilon}(m,\ell)p_{\kappa,0}(m,\ell)^2 - \p_\kappa p_{\kappa,0}(m,\ell)p_{\kappa_*,\epsilon}(m,\ell)^2}{p_{\kappa_*,\epsilon}(m,\ell)^2p_{\kappa_*,0}(m,\ell)^2}\rp|\notag\\
&\lesssim \fr{4\kappa_*^2 \epsilon m^2 \ell^2}{|p_{\kappa_*,\epsilon}(m,\ell))|^2} + \fr{|\p_\kappa p_{\kappa,0}(m,\ell)|}{|p_{\kappa,0}(m,\ell)|} \lp| \widehat{\mc{M}}(\kappa_*,\epsilon) - \widehat{\mc{M}}(\kappa_*,0) \rp|\notag\\
&\lesssim \epsilon.
\end{align}
Here we have used the fact that $|\fr{\p_\kappa p_{\kappa,0}(m,\ell)}{p_{\kappa,0}(m,\ell)}|$ is bounded in $(m,\ell)$ and the same scaling arguments used above for continuity of $\mc{M}$.  Combining the estimates for $\mathrm{(I)}$ and $\mathrm{(II)}$, we conclude the continuity of $\p_\kappa \mc{M}$ near $(\kappa_*,0),$ thus finishing the proof of the proposition.
\end{Proof}
\begin{Remark}\label{r:sm}
We note that $\mc{M}$ is readily seen to be smooth in $\epsilon$ and $\kappa$ as long as $\epsilon>0$. 
\end{Remark}
We can now conclude the continuity of $\mc{F} = w + \mc{M}(\kappa,\epsilon)\lp(w + f'(u_\rtr^*)w + \mc{R}(\kappa,\epsilon) + \mc{N}(w;\kappa,\epsilon)  \rp)$ in $\epsilon$, which shall allow us to apply the Implicit Function Theorem.

\begin{Corollary}\label{c:mf}
For $(\kappa,\epsilon)\sim (\kappa_*,0)$, the mapping $\mc{F}:X_\eta\times \R^2\rightarrow X_\eta$ is locally well-defined in a neighborhood of $(0;\kappa_*,0)$, smooth in $w$, $C^1$ in $\kappa$, and continuous in $\epsilon$.  Furthermore, the derivatives $\p_{w}\mc{F}$ and $\p_{\kappa}\mc{F}$ are also continuous in $\epsilon$.
\end{Corollary}
\begin{Proof}
We have that 
\begin{align}
\mc{F}(w;\kappa,\epsilon) &= \mc{M}(\kappa,\epsilon)\lp[ (\mc{L}(\kappa,\epsilon) w - w) + w + f'(u_\rtr^*)w + \mc{R}(\kappa,\epsilon)+\mc{N}(w;\kappa,\epsilon)  \rp]\notag\\
&= w + \mc{M}(\kappa,\epsilon)\lp[ w + f'(u_\rtr^*)w + \mc{R}(\kappa,\epsilon)+\mc{N}(w;\kappa,\epsilon)  \rp].\notag
\end{align}
We have already shown above that $\mc{R}(\kappa,\epsilon)$ and $\mc{N}(w,\kappa,\epsilon)$ are exponentially localized, taking values in $L^2_\eta$. A bootstrapping argument can then be used to obtain higher regularity of $u_\rtr^*$ in $\tau$ and $\zeta$. This along with the fact that $X_\eta$ is a Banach algebra, implies that $\mc{R}$ and $\mc{N}$ take values in $X_\eta$.  Hence we obtain that $\mc{F}$ is well-defined and smooth in $w$ near $w = 0$. 

Then, since
$$
\p_\kappa \Phi(\bk) = \p_2u_\mathrm{p}(\zeta+\tau;|\bk|)\fr{-2\kappa}{|\bk|^2} + \p_1u_\mathrm{p}(\kappa_*\zeta/\kappa+\tau;|\bk^*|)\fr{\kappa_*}{\kappa^2}\zeta,
$$
and
\begin{align}
&|u_\rtr^*(\kappa_*\zeta/\kappa,\tau) - u_\mathrm{p}(\kappa_*\zeta/\kappa + \tau;|\bk^*|)|\rightarrow 0\notag\\
 &|\p_\kappa u_\rtr^*- \p_\kappa u_\mathrm{p}(|\bk^*|)| = 
 \lp|\p_\xi u_\rtr^*(\kappa_*\zeta/\kappa,\tau)\fr{\kappa_*\zeta}{\kappa^2} - \p_1u_\mathrm{p}(\kappa^*\zeta/\kappa+\tau;|\bk^*|)\fr{\kappa_*\zeta}{\kappa^2}\rp|\rightarrow 0,
 \end{align}
  exponentially fast as $\zeta\rightarrow0$, one can readily show that the derivatives $\p_\kappa \mc{R}$ and $\p_\kappa \mc{N}$ are well defined as maps (in $\kappa$ and $w$ respectively) and continuous in $\kappa$ in a neighborhood of $(\kappa_*,0)$.   The regularity of $\mc{M}$ in $\kappa$ then implies that $\mc{F}$ is $C^1$ in $\kappa$ as well.
 We infer continuity of $\mc{F}$ in $\epsilon$ by combining Proposition \ref{p:m1} with the estimate
\begin{align}
\|\mc{F}(w;\kappa,\epsilon) - \mc{F}(w;\kappa,0)\|_{X_\eta} &\leq \|\mc{M}(\kappa,\epsilon) - \mc{M}(\kappa,0)\|_{X_\eta\rightarrow X_\eta} \| w +  f'(u_\rtr^*)w + \mc{R}(\kappa,\epsilon) +\mc{N}(w;\kappa,\epsilon) \|_{X_\eta}
\notag\\
&\quad\quad\quad+ \|\mc{M}(\kappa,0)\|_{X_\eta\rightarrow X_\eta} \| \mc{R}(\kappa,\epsilon) +\mc{N}(w;\kappa,\epsilon)  -  \mc{R}(\kappa,0) +\mc{N}(w;\kappa,0) \|_{X_\eta}.
\end{align}

In a similar manner, we also obtain that
\begin{align}
\p_w \mc{F}  &= I + \mc{M}(\kappa,\epsilon)\lp[ (1+f'(u_\rtr))I + \mc{R}(\kappa,\epsilon) +\mc{N}'(w;\kappa,\epsilon)  \rp],\notag\\
\p_{\kappa} \mc{F} &= \p_{\kappa}\mc{M}(\kappa,\epsilon)\lp[ (1+f'(u_\rtr))I + \mc{R}(\kappa,\epsilon)  + \mc{N}(w;\kappa,\epsilon)  \rp] + \mc{M}(\kappa,\epsilon) \lp[  \p_{\kappa}\mc{R}(\kappa,\epsilon) + \p_{\kappa}\mc{N}(w;\kappa,\epsilon)\rp],\notag
\end{align}
are continuous in $\epsilon$. 
\end{Proof}

\subsection{Proof of existence of oblique stripe formation}\label{ss:pfex}
To obtain the existence of oblique stripes we must study the linearization of the regularized mapping \eqref{e:mf} with respect to $(w,k_x)$ to obtain a family of solutions nearby $(w;\kappa,\epsilon) = (0;\kappa_*,0)$.  Hence we must consider the Fredholm properties of the linearization $\p_w\mc{F}\big|_{0,\kappa_*,0} = \mc{M}(\kappa_*,0)\circ\mb{L}: X_\eta\rightarrow X_\eta$.  
\begin{Lemma}\label{l:fredMF}
The operator $\mc{M}(k_x^*,0)\circ\mb{L}: X_\eta\rightarrow X_\eta$ is Fredholm with index $-1$ and one-dimensional cokernel.
\end{Lemma}
\begin{Proof}
 First note that $\mc{M}(\kappa_*,0)$ is bounded invertible as a map from $L^2_\eta\rightarrow X_\eta$ and hence is Fredholm with index 0.  Using Fredholm algebra and Lemma \ref{l:fc}, we then obtain that $\mc{M}(\kappa_*,0)\circ\mb{L}$ is Fredholm of index -1.
\end{Proof}

To define an adjoint of the linearization $\p_w\mc{F}\Big|_{0,\kappa_*,0}$ we first define a suitable adjoint for $\mc{M}(\kappa_*,0).$ A closed and densely defined adjoint can readily be defined for $\mc{M}^{-1}:X_\eta\subset L^2_\eta\rightarrow L^2_\eta$ as $(\mc{M}^{-1})^\mathrm{ad}:= \re^{\eta\la x\ra} \mc{L}(\kappa_*,\epsilon)\re^{-\eta\la x\ra} - \mathrm{id}$.
  Then, using for example \cite[Thm. III.5.30]{kato2013perturbation},  one can conclude the existence of $\mc{M}^\mathrm{ad}$ on $L^2_\eta$ and that $(\mc{M}^{-1})^\mathrm{ad} = (\mc{M}^\mathrm{ad})^{-1}$.  We thus have 
$$
\lp( \p_w\big|_{0,\kappa_*,0}\mc{F}\rp)^\mathrm{ad}:=(\mc{M} \mb{L})^\mathrm{ad} = \mb{L}^\mathrm{ad} \mc{M}^\mathrm{ad}.
$$ 
 Next, it is readily found that the cokernel of this operator is spanned by $v_* := (\mc{M}^\mathrm{ad})^{-1}e_*.$  We then have by Lemma \ref{l:fredMF},
 \begin{align}
 \la \p_{\kappa}\mc{F}, v_*\ra_{L^2_\eta}  &= \la \mc{M} \p_{\kappa}F + \p_{\kappa}\mc{M}F,(\mc{M}^{-1})^\mathrm{ad}e_*\ra_{L^2_\eta}  =  \la \p_{\kappa}F,e_*\ra_{L^2_\eta}  \neq0,
 \end{align}
where all derivatives and functions are evaluated at $(w,\kappa,\epsilon) = (0,\kappa_*,0)$.

A Fredholm bordering lemma (see for example \cite[\S 4.2]{sandstede2008relative}) then gives $[\p_{w}\mc{F},\p_{\kappa}\mc{F}] = [\mc{M}\circ\mb{L}, \mc{M}\circ\p_{k_x}F]:X_\eta\times \R\rightarrow X_\eta$ is Fredholm with index zero, has trivial kernel, and is thus invertible.  Combining this with Corollary \ref{c:mf}, we apply the Implicit Function theorem to obtain a continuous family of solutions $(w_\rtr(\epsilon),\kappa_\rtr(\epsilon))$ for all $\epsilon$ sufficiently small.  From this we obtain the existence of a continuous family of oblique stripes nearby the planar front $u_\rtr^*$,
\begin{equation}
u_\rtr(\epsilon) = u_\rtr^* + w_\rtr(\epsilon) + \chi(\Phi(\bk_\rtr(\epsilon)) - \Phi(\bk_*)
\end{equation}  
with wavenumber vector $\bk_\rtr(\epsilon) = (\kappa_\rtr(\epsilon),\sqrt{\epsilon})$.

\subsection{Differentiability and leading order expansion}\label{ss:boot}
Having proved the existence of a continuous family of solutions $(w_\rtr(\epsilon),k_x^\rtr(\epsilon))$ for $\epsilon$ near $0$, we now use a formal expansion and bootstrapping procedure to prove that this family is in fact differentiable at $\epsilon = 0$.  First we set 
\begin{multline}
\Delta\mc{L}(\kappa,\epsilon) =  \mc{L}(\bk) - \mc{L}(\bk_*) \\= (\kappa - \kappa_*)\lp[-(\kappa+\kappa_*)\p_\zeta^2 - (\kappa + \kappa_*)(\kappa^2 + \kappa_*^2)\p_\zeta^4 + c(\p_\zeta-\p_\tau)\rp] - \epsilon^2\p_\tau^4 - 2\epsilon (1+\kappa^2\p_\zeta^2)\p_\tau^2   .
\end{multline}
Then, momentarily omitting $\mc{M}$, we insert the formal ansatz
$$
w = \epsilon w_1 + \epsilon^2\tl w,\quad \kappa = \kappa_* + \epsilon \kappa_1 + \epsilon^2 \tl\kappa,
$$
into \eqref{e:mf}, 
\begin{align}
0 &= [\mc{L}(\bk^*) +f'(u_\rtr^*)+  \Delta\mc{L}(\kappa,\epsilon)](\epsilon w_1 + \epsilon^2 \tl w) \notag\\
&\qquad\qquad\qquad+ \mc{R}( \kappa_* + \epsilon \kappa_1 + \epsilon^2 \tl\kappa, \epsilon) + \mc{N}(  \epsilon w_1 + \epsilon^2\tl w,  \kappa_* + \epsilon \kappa_1 + \epsilon^2 \tl\kappa, \epsilon)\notag\\
&=\epsilon( \mb{L} w_1 + \mc{R}_1 \kappa_1 + \mc{R}_2) +  \epsilon^2[\Delta\mc{L}(\kappa,\epsilon)/\epsilon]( w_1 + \epsilon \tl w) + \epsilon^2(\mb{L}\tl w + \mc{R}_1 \tl \kappa) \notag\\
&\qquad\qquad\qquad + \tlR(\epsilon\kappa_1+\epsilon^2 \tl\kappa,\epsilon) + \tilde{\mc{N}}( \epsilon w_1 + \epsilon^2\tl w,  \epsilon \kappa_1 + \epsilon^2 \tl\kappa, \epsilon),\label{e:reg1}
\end{align}
where
\begin{align}
\tlR(K,\epsilon) &= \mc{R}(\kappa_*+K,\epsilon) - \mc{R}_1K - \mc{R}_2 \epsilon  \,\,=  \mc{O}\lp((|K| + |\epsilon|)^2\rp),\notag\\
\tlN(w,K,\epsilon)& = \mc{N}(w,\kappa_*+K,\epsilon) \,\, =  \mc{O}\lp(|w|(|w|+|K|+|\epsilon|)\rp), \notag
\end{align}
and 
\begin{align}
\mc{R}_1 &:= \p_{\kappa}\mc{R}(\kappa_*,0) = c\p_\tau u_\rtr^* + \mb{L}(\chi \p_{\kappa}\Phi(\bk_*)),\quad    \notag\\
\mc{R}_2 &:= \p_\epsilon\mc{R}(\kappa_*,0) = -2\p_\tau^2(1+(\kappa_*\p_\zeta)^2)u_\rtr^*+ \mb{L}(\chi \p_{\epsilon}\Phi(\bk_*)).   \notag
\end{align}
It this then readily seen that  $\tlR(\epsilon\kappa_1+\epsilon^2 \tl\kappa,\epsilon),\tilde{\mc{N}}( \epsilon w_1 + \epsilon^2\tl w,  \epsilon \kappa_2 + \epsilon^2 \tl\kappa, \epsilon)= \mc{O}(\epsilon^2)$.
Thus, we can solve \eqref{e:reg1} at $\mc{O}(\epsilon)$ by solving 
\begin{equation}\label{e:reg2}
 \mb{L} w_1 + \mc{R}_1 \kappa_1 = -\mc{R}_2.
\end{equation}
By bootstrapping to higher regularity in space and time for $u_\rtr^*$ one obtains that $\mc{R}_2\in L^2_\eta$.  As in Section \ref{ss:pfex}, this equation is affine linear and the augmented operator $[\mb{L}, \mc{R}_1]: X_\eta\times\R \rightarrow L^2_\eta$ is well-defined and invertible by Lemma \ref{l:fredMF}. 
Indeed, integration by parts using the exponential localization of $e_*$ gives,
$$
\la \mb{L}(\chi \p_{\kappa}\Phi(\bk_*),e_*\ra_{L^2_\eta} = 0.
$$
Then since $\la \p_\tau u_\rtr^*,e_*\ra_{L^2_\eta} = 1,$ we obtain
$$
\la \mc{R}_1,e_*\ra_{L^2_\eta} = c \neq 0. 
$$
Thus we can solve, obtaining
$$
(w_1,\kappa_1) = -[\mb{L}, \mc{R}_1]^{-1}\mc{R}_2 \in X_\eta\times\R.
$$
Inserting this solution into \eqref{e:reg1}, and dividing by $\epsilon^2$ we then obtain an equation for the residual $(\tl w,\tl \kappa)$,
\begin{align}
0&= \mc{M}(\epsilon \kappa_1 + \epsilon^2\tl\kappa,\epsilon)\Bigg[ (\mb{L}\tl w + \mc{R}_1 \tl \kappa + \Delta\mc{L}(\kappa,\epsilon)/\epsilon)(w_1+\epsilon \tl w) + \fr{\tlR}{\epsilon^2} + \fr{\tlN}{\epsilon^2}
\Bigg].\label{e:reg3}
\end{align}
We then notice that at $\epsilon = 0$ this equation collapses to
\begin{align}
0 &= \mc{M}(k_x^*,0)\Bigg[ \mb{L}\tl w + \mc{R}_1 \tl \kappa + \kappa_1(2\kappa_*\p_\zeta^2 + 4\kappa_*^3\p_\zeta^4 + c(\p_\zeta+\p_\tau))w_1 + 2\p_\tau^2(1+\kappa_*\p_\zeta^2)w_1\notag\\
&\quad\quad\quad+ \mc{R}_{11}\kappa_1^2 + 2 \mc{R}_{12} \kappa_1 + \mc{R}_{22} + f''(u_*)[w_1^2 + 2\chi w_1(\kappa_1\p_{\kappa}\Phi + \p_{\epsilon}\Phi)]  \Bigg],
\end{align}
because $\p_2\tlN = \p_3\tlN = \p_{22}\tlN = \p_{33}\tlN = 0$.  This equation is affine linear in $(\tl w,\tl\kappa)$ and we once again obtain a solution $(\tl w_0,\tl\kappa_0)\in X_\eta$ by inverting the augmented operator $[\mb{L},\mc{R}_1]$.
Here $\mc{R}_{ij} = \p_{ij}\mc{R}(\kappa_*,0)$ for $i,j = 1,2$.  Since the linearization in $(\tl w,\tl \kappa)$ of this equation at $\epsilon = 0$ is the same as in \eqref{e:mf} we can apply the Implicit Function theorem to \eqref{e:reg3} just as in the previous section to obtain the existence of a continuous family of solutions $(\tl w,\tl\kappa)(\epsilon)\in X_\eta\times\R$ with $(\tl w,\tl\kappa)(0) = (\tl w_0,\tl\kappa_0).$  

By the uniqueness of the Implicit Function theorem result, and the fact that
$$
\|(\epsilon w_1 + \epsilon^2\tl w(\epsilon), \epsilon \kappa_1 + \epsilon^2 \kappa(\epsilon)) - (\epsilon w_1 , \epsilon \kappa_1 )\|_{X_\eta\times\R} = \epsilon^2\|( \tl w(\epsilon),  \tl\kappa(\epsilon)) \|_{X_\eta\times\R}  = \mc{O}(\epsilon^2),
$$
we obtain that the solution family and wavenumber are differentiable in $\epsilon$ at $\epsilon = 0$.  Furthermore, the $\mc{O}(\epsilon)$-equation \eqref{e:reg2} can be used to obtain the leading order wavenumber correction, $\kappa_1$, in $\epsilon$.  Namely, one projects \eqref{e:reg2} onto the cokernel of $\mb{L}$, obtaining
\begin{equation}\label{e:coef}
\kappa_1 = - \fr{b_y}{b_x},\qquad b_x:={\la \mc{R}_1, e_*\ra} = c,\,\, b_y :=\la\mc{R}_2,e_*\ra =\la -2\p_\tau^2(1+\kappa_*^2\p_\zeta^2) u_\rtr^*,e_*\ra  .
\end{equation}
so that the selected wavenumber has the leading order expansion
\begin{equation}
k_x(k_y) = \kappa(k_y^2) = k_x^*  - \fr{b_y}{b_x}\epsilon + \mc{O}(\epsilon^2).
\end{equation}
This completes the proof of Theorem \ref{t:2}.

\begin{Remark}\label{r:kexpn}
From this result an expansion in $k_y$ for the modulus, or total strain, of the striped pattern can readily be obtained as 
\begin{equation}
|\bk|^2 = k_x(k_y)^2+k_y^2 =  (k_x^*)^2+ (1-2\fr{b_y}{b_x} k_x^*) k_y^2 + \mc{O}(k_y^4).
\end{equation}
\end{Remark}

\section{Parallel stripe formation near onset --- proof of Theorem \ref{t:1}}\label{s:on}

We construct an example of a planar trigger front that satisfies the hypotheses of our main result, Theorem \ref{t:2}. We therefore solve \eqref{e:par}, arising from the Swift-Hohenberg equation through an ansatz $u=u(\omega t,k_x\zeta)$, with $\omega=ck_x$. Our focus here is on small-amplitude solutions with small speed $c$. In this parameter regime, it turns out that a slightly different choice of coordinates is more convenient. Therefore, we try to find $u(x,t) = W(x,x-ct),$ with $W(x,\xi) = W(x+(2\pi/k),\xi)$, corresponding to a shear transformation from \eqref{e:par} with $\tau\to \tau+\xi$ and a scaling by $k_x$. 

We obtain existence through a center-manifold based approach used to construct front solutions in various settings \cite{eckmann1991propagating,doelman2003propagation,huaruagucs1999bifurcating}.  In such works, fronts  are studied as solutions of (ill-posed) dynamical systems with $\xi$ as the time-like evolutionary variable,  near the onset of Turing instability at $\mu_0 = 0$, making the parameter scaling
$$
\mu(x-ct) = -\epsilon^2\mathrm{sgn}(x-ct),\quad c = \epsilon \tl c, \quad 0<\epsilon\ll1.
$$
After a Fourier decomposition in $x$,  the linearized system about the origin is found to possess infinitely many center eigenvalues for $\epsilon = 0$.  When $\epsilon$ becomes positive, all but finitely many of these eigenvalues move away from the imaginary axis with $\mc{O}(\epsilon^{1/2})$ rate, leaving a finite set of isolated eigenvalues in an $\mc{O}(\epsilon)$-neighborhood of $\ri\R$.  One then performs a center-manifold reduction about the eigenspace associated with these $\mc{O}(\epsilon)$-eigenvalues to obtain a finite-dimensional phase-space which organizes the dynamics. 
Within this finite-dimensional center-manifold, the spatial dynamics possess equilibria for $\mu\equiv \epsilon^2$, that is, for $\xi<0$, corresponding to spatially periodic solutions with wavenumber $k$, with explicit leading order form
$$
u_\mathrm{p}(k x;k) = \epsilon\sqrt{4(1-4\gamma^2)/3} \cos(kx) +\mc{O}(\epsilon^2),\qquad k=1+\gamma \epsilon.
$$

Pattern-forming fronts can then be obtained by finding intersections between the center-unstable manifold of the periodic orbit in the $\xi<0$-dynamics with the stable manifold of the origin in the $\xi>0$-dynamics. These invariant manifolds can be foliated by fibers with base-points on the respective center-manifolds.  We shall see that intersections of these invariant manifolds can be found to leading order by simply overlaying the dynamics of the center-manifolds and finding the desired intersections.    

To begin we collect information about the $\mu\equiv\pm\epsilon^2$ dynamics separately, with the two equations
\begin{equation}\label{e:pm}
u_t = -(1+\p_x^2)^2 u + \epsilon^2\tl \mu^{\rrl} u - u^3,
\end{equation}
where $\tl \mu^{\rrl} = \mp1$ corresponding to $x-ct\gtrless0$ respectively. We use the ``${\rrl}$" notation to study both equations at once when possible. 

\paragraph{Center-manifold approach.}
Once again, we adapt the approach of \cite{eckmann1991propagating}, studying solutions of the form $$u(x,t) = W(x,x-ct),\quad W(x,\cdot) = W(x+2\pi/k,\cdot)$$ so that $W$ has the Fourier decomposition $W(x,\xi) = \sum_{n\in \Z} W_n(\xi) \re^{-\ri n k x}$. By decomposing \eqref{e:pm} into an infinite set of finite dimensional systems for each $W_n$, we are able to study spectra and perform a center-manifold reduction in each of the $\tl \mu^{\rrl}$ phase portraits.  The $\tl \mu^\mathrm{l}$ portrait will have a circle of equilibria which correspond to the periodic wave trains $u_\mathrm{p}$ of wavenumber $k$.  Substituting the decomposition of $u$ into \eqref{e:pm}, we obtain the equation
\begin{equation}\label{e:fft}
\lp[\epsilon^2 \tl \mu^{\rrl} + \epsilon \tl c \p_\xi - (1 + (-\ri  k n + \p_\xi)^2)^2\rp] W_n(\xi) = \sum_{p+q+r = n}W_p(\xi)W_q(\xi)W_r(\xi).
\end{equation}
 Then setting $X = (X_n)_{n\in \mb{Z}}$ with $X_n = (W_n,\p_\xi W_n,\p_{\xi}^2W_n,\p_\xi^3 W_n)^T$, we obtain the first-order system
 \begin{equation}\label{e:fs}
 \p_\xi X_n = M_n^{\rrl} X_n + F_n(X),
 \end{equation}
for each Fourier mode with 
$$
M_n^\rrl = \left(\begin{array}{cccc}0 & 1 & 0 & 0 \\0 & 0 & 1 & 0 \\0 & 0 & 0 & 1 \\A^\rrl & B & C & D\end{array}\right), 
\quad
F_n(X) = (0,0,0, \sum_{p+q+r = n} X_{p,0}X_{q,0}X_{r,0})^T,
$$
and 
$$
A^{\rrl} = -(1 - (kn)^2)^2 + \epsilon^2\tl \mu^{\rrl},\quad B = 4\ri kn(1 - (kn)^2) + c,\quad C = 6(kn)^2 - 2,\quad D = 4\ri kn.
$$
The characteristic polynomial for each $M_n^{\rrl}$ is found to be
$$
p_n^{\rrl}(\nu) = (\nu - \ri(kn + 1))^2(\nu - \ri(kn - 1))^2 - \tl c\epsilon \nu - \tl \mu^{\rrl} \epsilon^2,
$$
so that $M_n^{\rrl}$ has a pair of geometrically simple and algebraically double eigenvalues at $\nu = \ri(kn \pm 1)$ for $\epsilon = 0$.
As in \cite{eckmann1991propagating} we let $k = 1+\epsilon \tl \gamma$ and find that for $0<\epsilon\ll1$, all eigenvalues $\nu$   are at least $\mc{O}(\epsilon^{1/2})$ distance away  from $\ri\R$ except for the pairs 
$$
\nu_{n,\pm}^{\rrl} = \epsilon (\chi^{\rrl}_\pm + \ri \tl \gamma) + \mc{O}(\epsilon^2),\quad \chi^{\rrl}_{\pm} = \fr{-\tl c \pm \Delta^{\rrl}}{8},\quad \Delta^{\rrl} = \sqrt{\tl c^2 - 16(\tl \mu^{\rrl} +\ri \tl c\tl \gamma)},
$$
in the $n = \pm1$ subspaces, with corresponding eigenvectors
$
\phi^{\rrl}_\pm = (1,\nu_{n,\pm}^{\rrl},(\nu_{n,\pm}^{\rrl} )^2,(\nu_{n,\pm}^{\rrl} )^3)^T.
$
Also let $\psi_\pm^{\rrl}$ be the corresponding adjoint eigenvectors.

Following, \cite{eckmann1991propagating} we let $\mc{E}_0 = \bigoplus_{n=0}^\infty\C^4$ and $X_{nj}$ be the $j$-th component of $X_n$ for $j = 0,...3$ and $\mc{E} := \{X\in \mc{E}_0\,|\, X_{0j} \in \R\}.$   Furthermore, the inner product 
$$
\la X, Y\ra_{s} = \sum_{n=0}^\infty(1+n^2)^s\la X_n,Y_n\ra _{\C^4},\quad s\geq 0,
$$
makes $\mc{E}_0$ a complex Hilbert space, which we denote by $H$.  We let $\Phi_\pm^\rrl \in \mc{E}$ be the vector representation of the eigenvectors of $M_1^\rrl$ found above,
$$
(\Phi_\pm^\rrl)_1 = \phi_\pm^\rrl,\quad (\Phi_\pm^\rrl)_n = 0, n\neq 1,
$$
and define $\Psi_{\pm}$  in a similar way for the adjoint eigenvectors $\psi_{\pm}.$  We can then define a spectral projection $P^{c}:\mc{E}_0\rightarrow E^\mathrm{c}:=\mathrm{span}\{\Phi_+,\Phi_-\}$ as
\begin{equation}\label{e:proj}
P^\mathrm{c}_\rrl X = c_+ \la \Psi^\rrl_+, X\ra_s \Phi^\rrl_+ + c_-\la \Psi^\rrl_-,X\ra_s \Phi^\rrl_-, 
\end{equation}
with the normalization constants
$$
c_\pm^\rrl =  \fr{\mp1}{\epsilon \Delta_\rrl}\lp( 1+ \mc{O}(\epsilon)\rp).
$$
Since $W_{-n} = \overline{W}_{n}$, we  only need  to study the center directions in the $n = 1$ Fourier subspace.  Thus we let $\nu_\pm^{\rrl} = \nu_{1,\pm}^{\rrl}$ for the rest of the proof.

 We can then apply Theorem A.1 of \cite{eckmann1991propagating} or Theorem 6.3 of \cite{huaruagucs1999bifurcating} to obtain a local center manifold $W_\rrl^\mathrm{c}(0)$ of the system \eqref{e:fs} described by the graph $(w,h_\rrl(w))$ with 
 $$
  h_\rrl:E_\rrl^\mathrm{c} \rightarrow (E_\rrl^\mathrm{c})^{\perp},\quad h_\rrl(w) = \mc{O}(\|w\|^3),
 $$ 
 where the orthogonal complement is taken in $\mc{H}$. Note that the first cited theorem gives $C^1$-smoothness while the latter gives $C^m$-smoothness for $m>1$ with $\epsilon$ sufficiently small. We note that by the construction of $P^\mathrm{c}_\rrl$, the subspace $\mathrm{Rg}\, (1-P^\mathrm{c}_\rrl)$ consists of the hyperbolic, or non-center, eigenspaces, $E^{\rs,\ru}_\rrl$, of the linear operators $M^\rrl_n$. In the coordinates  $w = a_+ \Phi_+ + a_-\Phi_-$, the equation on the center manifold takes the form
\begin{align}\label{e:xsys}
\fr{da_+}{d\xi}&= \nu_{+}^{\rrl} a_+ - 3 c_+ (a_+ + a_-)|a_+ + a_-|^2 + \mc{O}(|a_++a_-|^4),\notag\\
\fr{da_-}{d\xi}&= \nu_{-}^{\rrl}a_- - 3c_-(a_+ + a_-)|a_+ + a_-|^2 + \mc{O}(|a_++a_-|^4).
\end{align}

By rescaling time $\zeta = \epsilon \xi$ and applying the linear change of coordinates
\begin{align}
\left(\begin{array}{c}q \\ p \end{array}\right) &= \fr{1}{\epsilon}\left(\begin{array}{cc}1& 0 \\-\fr{\tl c}{8}+\tl \gamma \ri & \fr{\Delta^{\rrl}}{8}\end{array}\right)\left(\begin{array}{cc}1 & 1 \\1 & -1\end{array}\right)\left(\begin{array}{c}a_+\\ a_- \end{array}\right)  \notag\\
&= \fr{1}{\epsilon}\left(\begin{array}{cc}1 & 1 \\ \chi_+^{\rrl}+\ri\tl \gamma & \chi_-^{\rrl}-+\ri\tl \gamma\end{array}\right)\left(\begin{array}{c}a_+\\ a_- \end{array}\right)=:B^{\rrl}(\epsilon) \left(\begin{array}{c}a_+\\ a_- \end{array}\right)\label{e:Beps}
\end{align}
the system \eqref{e:xsys} can be put into normal form
\begin{align}\label{e:red}
\fr{dq}{d\zeta}&= p+ \mc{O}(\epsilon),\notag\\
\fr{dp}{d\zeta}& = \fr{1}{4} \lp( -\tl \mu^{\rrl} q - \tl c p + 3 q|q|^2\rp) + \tl \gamma(2\ri p + \tl \gamma q)  + \mc{O}(\epsilon).
\end{align}
 Note that this is also the equation obtained if one performs a multiple scale expansion, inserting the ansatz $u = \epsilon A(\epsilon x, \epsilon^2\tau) \re^{\ri k x} + \mathrm{c.c.}$ into the original equation with parameters scaled, taking leading order terms in $\epsilon$ and then transforming into a moving frame.  We now collect some facts about the leading order system.
\begin{Proposition}
The following hold for the leading order system \eqref{e:red} with $\epsilon =0$, $\tl \gamma=0$ and $0<4-\tl c\ll1$.  For $\tl \mu^\mathrm{r} = -1$,
\begin{itemize} 
\item The origin $(0,0)$ is a hyperbolic equilibrium. When formulated as a real system by decomposing into real and imaginary parts, it has double eigenvalues $\chi_\pm^\mathrm{r}$.
\end{itemize}
 For $\tl \mu^\mathrm{l} = 1$,
\begin{itemize}
\item The origin $(0,0)$ is a stable equilibrium. When formulated as a real system, it is a stable spiral with double eigenvalues $\chi_\pm^\mathrm{l}$.
\item The point $\mc{P}_0=(q_0,p_0)= 1/\sqrt{3}$ is an equilibrium with eigenvalues $0, -\tl c, \fr{-\tl c \pm \sqrt{\tl c^2+32}}{8}$. 
\item Due to the phase-invariance of the system under rotations $(q,p)\mapsto \re^{\ri\theta}(q,p)$, each point on the circle  $\mc{P}= \{(\re^{\ri\theta}q_0,\re^{\ri\theta}p_0)\,|\,\theta\in [0,2\pi)\}$ is also an equilibrium with the same stability properties.
\item There exists a one-parameter family of heteroclinic orbits $\mc{H}$ connecting the equilibria in $\mc{P}$ to the origin $(0,0)$. 
\end{itemize}
Furthermore, these properties persist for $\tl \gamma \sim 0$ except for the multiplicities of the eigenvalues and the equilibria position $|q_*|^2 = (\tl \mu - 4\tl \gamma)/3$.
\end{Proposition}
We denote the stable manifold about $(0,0)$ with $\tl \mu^\mathrm{r}$ as $\tl W_\mathrm{r}^\rs(0)$ and the center-unstable manifold of the circle of equilibria for $\mu = \tl\mu^\mathrm{l}$ as $\tl W^\rcu_\mathrm{l}(\mc{P})$. Finally, we note this family forms a normally hyperbolic invariant manifold and thus persists under $\mc{O}(\epsilon)$-perturbations as a one dimensional family of equilibria in \eqref{e:red}; see \cite[\S B]{eckmann1991propagating}.

\paragraph{Invariant Foliations.}
 Since $E_c^\mathrm{r}$ and $E_c^\mathrm{l}$ do not span the same space for $\epsilon>0,$ the corresponding invariant manifolds $W^c_\rrl$ do not necessarily intersect away from the origin. To see this formally restrict to the $n=1$ Fourier space, then $W^c_\mathrm{r}$ and $W^c_\mathrm{l}$, each having dimension four in the decomplexification, $\R^8$, of $\C^4$, generically have zero-dimensional intersection.  Thus, we can not directly seek intersections of the submanifolds $\tl W_\mathrm{l}^\rcu(\mc{P})$ and $\tl W_\mathrm{r}^\rs(0)$ contained in the respective center-manifolds, and must organize the local hyperbolic dynamics nearby. 

Hence we construct a pattern-forming front by finding intersections between the full stable manifold, $W_\mathrm{r}^\rs(0)$, of the origin and the center-unstable manifold, $ W_\mathrm{l}^\rcu(\mc{P})$, of the family of equilibria $\mc{P}$. Using standard results on foliations of center-manifolds one can construct strong-stable and strong-unstable foliations locally around each center manifold $W^\mathrm{c}_\rrl(0)$.  Then we find an intersection by projecting $W^\rs_\mathrm{r}(0)$, the union of $\tl W_\mathrm{r}^\rs(0)$ and its strong-stable foliation, onto $W_\mathrm{l}^\mathrm{c}(0)$ along its own strong-unstable foliation. We then show that to leading order, this is equivalent to overlaying the center-manifold dynamics in \eqref{e:red} for $\tl\mu^\mathrm{r}$ on top of those for $\tl \mu^\mathrm{l}$. Hence trigger fronts can be found by finding intersections of the submanifolds $\tl W_\mathrm{l}^\rcu(\mc{P})$ and $\tl W_\mathrm{r}^\rs(0)$.

In more detail, general results on invariant foliations \cite{fenichel1977asymptotic} give that the center-manifolds $W_\mathrm{r}^\mathrm{c}(0)$ and $W_\mathrm{l}^\mathrm{c}(0)$ possess local strong-stable and strong-unstable $C^m$-smooth foliations
corresponding to the stable and unstable eigenvalues at least $\mc{O}(\epsilon^{1/2})$ distance away from the imaginary axis, $$
\mc{F}^{\mathrm{ss}/\mathrm{uu}}_\mathrm{r} = \bigcup_{w\in W_\mathrm{r}^\mathrm{c}(0)} \mc{F}^{\mathrm{ss}/\mathrm{uu}}_{\mathrm{r}, w},\quad
\mc{F}^{\mathrm{ss}/\mathrm{uu}}_\mathrm{l} = \bigcup_{w\in W_\mathrm{l}^\mathrm{c}(0)} \mc{F}^{\mathrm{ss}/\mathrm{uu}}_{\mathrm{l}, w}
$$
 with $C^m$-fibers $\mc{F}^j_{\rrl,w}$ diffeomorphic to $E^j_\rrl$ and $C^{m-1}$-smooth dependence on the base-point $w\in W_\rrl^c(0)$. Furthermore, locally within the stable and unstable foliations, there exists a forward, repectively backward evolution $\Phi_\xi$ of \eqref{e:fs}, leaving these foliations invariant in the sense that
 $$
 \Phi_\xi(\mc{F}^{j}_{\rrl,w})\subset \mc{F}^j_{\rrl,\Phi_\xi(w)},
 $$ for $j = \mathrm{ss}$ with $\xi>0$, and for $j=\mathrm{uu}$ with $\xi<0$.   In particular each fiber $\mc{F}^j_{\rrl,w}$ can be written as a graph $v_\rrl^j + g_\rrl^j(v_\rrl^j,w_\rrl)$ with smooth functions $g_\rrl^j:E_\rrl^j\times E_\rrl^\mathrm{c}\rightarrow E_\rrl^\mathrm{c}$.    Also note that for $w\in \mc{P}$,
 \begin{align}
 W^\mathrm{uu}_\mathrm{l}(w) &=\mc{F}^\mathrm{uu}_ {\mathrm{l},w},\quad 
 W^\rcu_\mathrm{l}(\mc{P}) = \bigcup_{w\in \tl W_\mathrm{l}^\rcu(\mc{P})}\mc{F}^\mathrm{uu}_ {\mathrm{l},w}, \notag\\ 
 W^{\rss}_\mathrm{r}(0) &= \mc{F}^\mathrm{ss}_ {\mathrm{r},0},\quad
 W^\rs_\mathrm{r}(0) = \bigcup_{w\in \tl W_\mathrm{r}^\rs(0)} \mc{F}^\mathrm{ss}_ {\mathrm{r},w}.\notag
 \end{align}

 \begin{figure}[h!]\centering
\includegraphics[width=.7\textwidth]{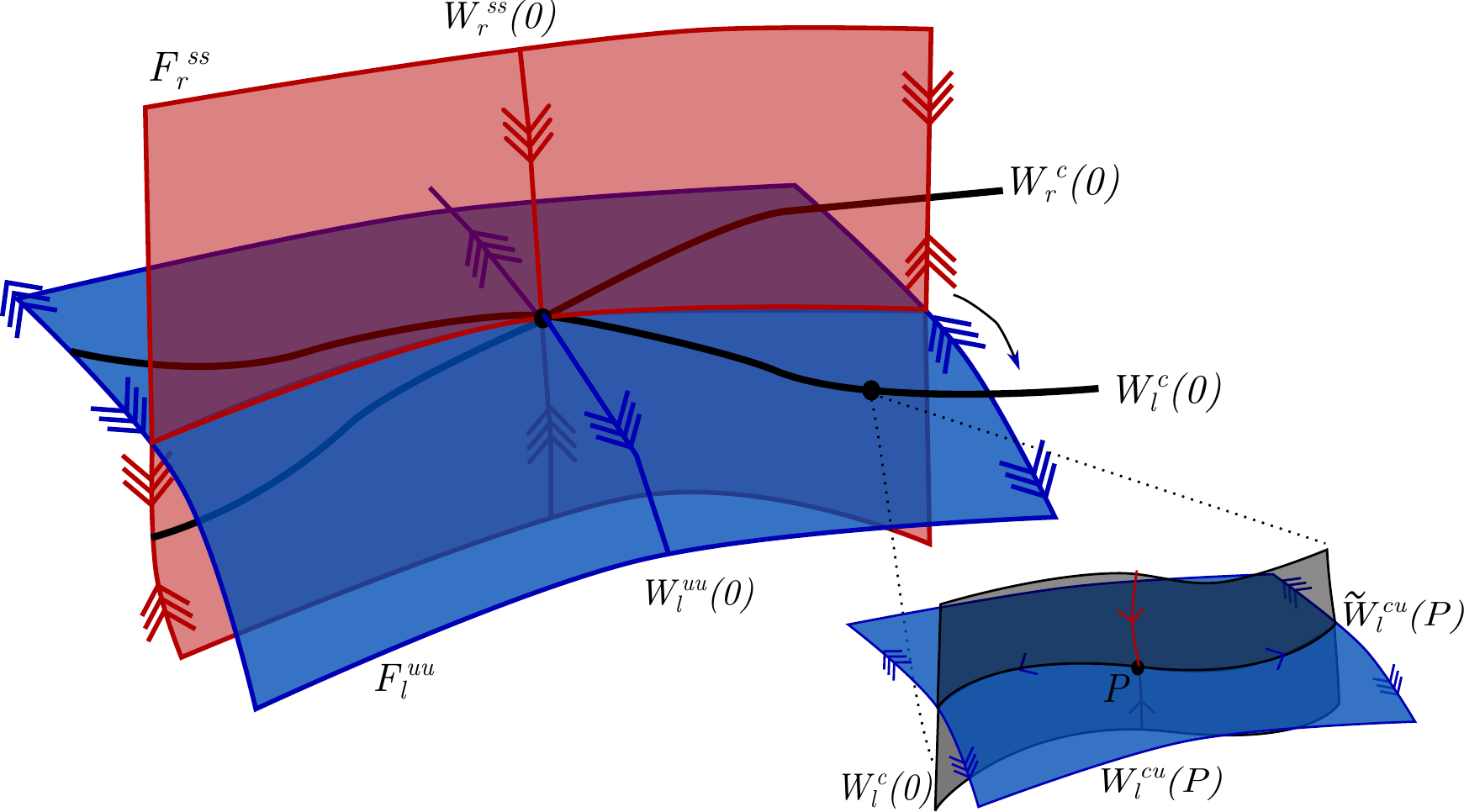}
\caption{Strong-stable (red) and Strong-unstable (blue) foliations around the center manifolds $W_\mathrm{\rrl}^c(0)$ (black curves). Inset depicts the strong-unstable foliation of the center manifold dynamics around the family of periodics $\mc{P}$ (represented by a point).}
\label{f:1}
\end{figure}

Next observe that in the $n=1$ Fourier space, $\mc{F}^{\mathrm{ss}}_\mathrm{r}$ and $\mc{F}^\mathrm{uu}_\mathrm{l}$ are each real six-dimensional and hence generically have a real four-dimensional intersection near the origin. Since this is the subspace containing the center directions, we expect that the full intersection $W_\mathrm{r}^\mathrm{cu}(0)\cap W_\mathrm{l}^\mathrm{cs}(0)$ will be real four-dimensional as well.  Using the graph descriptions of the invariant manifolds and foliations given above, this intersection is given by roots the equation
\begin{align}
\mc{T}&:E^\mathrm{c}_\mathrm{r}\times E^\mathrm{c}_\mathrm{l}\times E^\rs_\mathrm{r} \times E^\ru_\mathrm{l}\rightarrow \mc{E}_0\notag\\
&(w_\mathrm{r},w_\mathrm{l},v_\mathrm{r}^\rs,v_\mathrm{l}^\ru)\mapsto 
w_\mathrm{l} - w_\mathrm{r} + v_\mathrm{l}^\ru - v_\mathrm{r}^\rs 
+ h_\mathrm{l}(w_\mathrm{l}) - h_\mathrm{r}(w_\mathrm{r}) 
+ g_\mathrm{l}^\ru(v_\mathrm{l}^\ru,w_\mathrm{l}) - g_\mathrm{r}^\rs(v_\mathrm{r}^\rs,w_\mathrm{r}).
\end{align}
We then search for roots of $\mc{T}$ near by the origin where $\mc{T}(0,0,0,0) = 0$.

\begin{Proposition}\label{p:fol}
Fix $\epsilon>0$ sufficiently small. Then for all sufficiently small $w_\mathrm{r}\in E^\mathrm{c}_\mathrm{r}$, there exists a two-parameter family of solutions $(w_\mathrm{l},v_\mathrm{r}^\rs,v_\mathrm{l}^\ru)_*(w_\mathrm{r})$ of $\mc{T} = 0$.
 Furthermore, the solution $w_{\mathrm{l},*}(w_\mathrm{r})$ satisfies
\begin{equation}
w_{\mathrm{l},*}(w_\mathrm{r}) = P^\mathrm{c}_\mathrm{l}w_\mathrm{r} + \mc{O}(||w_\mathrm{r}||^2 +\epsilon^2),
\end{equation}
with the projection $P_\mathrm{l}^\mathrm{c}$ defined in \eqref{e:proj} above.
\end{Proposition}
\begin{Proof}
Given the smoothness of the foliation, we may straighten out the fibers via a smooth change of coordinates and study the map:
\begin{align}   
\tilde{\mc{T}}(w_\mathrm{r},w_\mathrm{l},v_\mathrm{r}^\rs,v_\mathrm{l}^\ru)=
w_\mathrm{l} - w_\mathrm{r} + v_\mathrm{l}^\ru - v_\mathrm{r}^\rs 
+ h_\mathrm{l}^\mathrm{c}(w_\mathrm{l}) - h_\mathrm{r}^\mathrm{c}(w_\mathrm{r} ).
\end{align}
Any corrections to the final solution $w_\mathrm{l}(w_\mathrm{r})$ from this coordinate change will come at higher order. We then prove the proposition using a Lyapunov-Schmidt reduction near the trivial solution $\tl T(0,0,0,0) =0$.  We define $P^\perp$ to be the projection in $\mc{E}_0$ onto $E^\rs_\mathrm{r}+ E^\ru_\mathrm{l}$ along $E_\mathrm{l}^\mathrm{c}$, and denote $P^\mathrm{c} = 1 - P^\perp$ as it's complement.  Due to the fact that $\mu_0^\rrl$ corrections come in at $\mc{O}(\epsilon^2)$  in the hyperbolic eigenspaces we have that $P^\perp = P_\mathrm{l}^\mathrm{c}+\mc{O}(\epsilon^2)$ in operator norm.  We then decompose the equation $\tilde{\mc{T}} = 0$, first solving 
$$
0=P^\perp\tilde{\mc{T}} = v_\mathrm{l}^\ru - v_\mathrm{r}^\rs + P^\perp\lp( h_\mathrm{l}(w_\mathrm{l}) - h_\mathrm{r}(w_\mathrm{r}) - w_\mathrm{r}  \rp),
$$
for $(v_\mathrm{l}^\ru,v_\mathrm{r}^\rs)$ in terms of $(w_\mathrm{r},w_\mathrm{l})$, as the linearization of this equation has $D_{(v_\mathrm{l}^\ru,v_\mathrm{r}^\rs)}P^\perp \tilde{\mc{T}} = \iota_\perp,$ where
$$
\iota_\perp:E^\rs_\mathrm{r}\times E^\ru_\mathrm{l}\rightarrow\mc{E}_0,\quad \iota_\perp(v,w) = v - w,
$$
is the joint canonical embedding which is invertible on its range $E^\rs_\mathrm{r}+ E^\ru_\mathrm{l}$.  We then solve the complimentary equation 
$$
0 = P^\mathrm{c}\tilde{\mc{T}} = w_\mathrm{l} + P^\mathrm{c}\lp( h_\mathrm{l}(w_\mathrm{l}) - w_\mathrm{r}  - h_\mathrm{r}(w_\mathrm{r}) \rp),
$$
which we can then readily solve for $w_\mathrm{l}$ in terms of $w_\mathrm{r}$.  The expansion above then follows from the fact that $ P^\mathrm{c} = P_\mathrm{l}^\mathrm{c} + \mc{O}(\epsilon)$ in the operator norm.
\end{Proof}

The desired trigger fronts are then given to leading order by projecting $\tl W^\rs_\mathrm{r}(0)$ onto $W^\mathrm{c}_\mathrm{l}(0)$ and finding intersections with $\tl W_\mathrm{l}^\rcu(\mc{P}).$ In the normal-form coordinates $\{q_\mathrm{r},p_\mathrm{r}\}$ and $\{q_\mathrm{l},p_\mathrm{l}\}$ the projection $P_\mathrm{l}^\mathrm{c}$ has the form $I_2 + \mc{O}(\epsilon)$, with $I_2$ the identity matrix on $\C^2$.  This can be found by a straightforward calculation using the transformations $B^\rrl(\epsilon)$, defined in \eqref{e:Beps} above, and the explicit form of the projection $P_\mathrm{l}^\mathrm{c}$ in \eqref{e:proj}. We can now prove the following theorem:
\begin{Theorem}\label{t:on-ex}(Existence)
For all $\epsilon>0$ sufficiently small, and $0<4-\tl c\ll1$, there exists a one-parameter family of planar trigger front solutions, $u^*_\epsilon(x-ct, t;\tl c)$, with asymptotic spatial wavenumber $k(\tl c) = 1 + \mc{O}(\epsilon)$, which are $2\pi/(ck)$-periodic in their second argument, and satisfy
\begin{align}
|u^*_\epsilon(\xi,t;\epsilon, \tl c) - u_\mathrm{p}(k \xi - ck t ; \tl c)| \rightarrow 0,\quad &\xi\rightarrow-\infty\notag\\
|u^*_\epsilon(\xi,t;\epsilon, \tl c)|\rightarrow 0,\quad &\xi\rightarrow+\infty.\notag
\end{align}
\end{Theorem}

\begin{Proof}
We begin by overlaying the two versions of \eqref{e:red} and studying the non-autonomous system
\begin{align}\label{e:red1}
\fr{dq}{d\zeta}&= p,\notag\\
\fr{dp}{d\zeta}& = \fr{1}{4} \lp( -\mathrm{sgn}(-\zeta) q - \tl c p + 3 q|q|^2\rp) + \tl \gamma(2\ri p + \tl \gamma q) ,\quad (p,q)\in \C^2.
\end{align}

 For $\gamma = 0$, an intersection of $\tl W_\mathrm{r}^\rs(0)$ with $\tl W_\mathrm{l}^\rcu(\mc{P}_0)$ can be found in the leading order system \eqref{e:red1} with $(q,p)\in \R^2$.  Indeed for any $0<4-\tl c\ll1$, the $\zeta<0$ phase portrait has a real heteroclinic orbit in the family $\mc{H}$ which connects the equilibria $\mc{P}_0=(q_0,p_0):=(\fr{1}{\sqrt{3}},0)^T$ to the origin $(0,0)$.  Since this heteroclinic spirals into the origin (as opposed to approaching it monotonically) it must intersect the tangent space $T_0 \tl W_\mathrm{r}^\rs(0)\cap \R^2$ near the origin, implying it must intersect $\tl W^\rs_\mathrm{r}(0)$ as well. Then setting this intersection point as an initial condition for \eqref{e:red1} at $\zeta = 0$ we obtain a heteroclinic front solution $(q_*(\xi),p_*(\xi))$ which converges to the desired equilibria in forwards and backwards time.  Furthermore, due to the rotation symmetry we obtain a one-parameter family of intersections of $\tl W_\mathrm{l}^\rcu(\mc{P})$ and $\tl W_\mathrm{r}^\rs(0).$
 
  \begin{figure}[h!]\centering
\includegraphics[width=.63\textwidth]{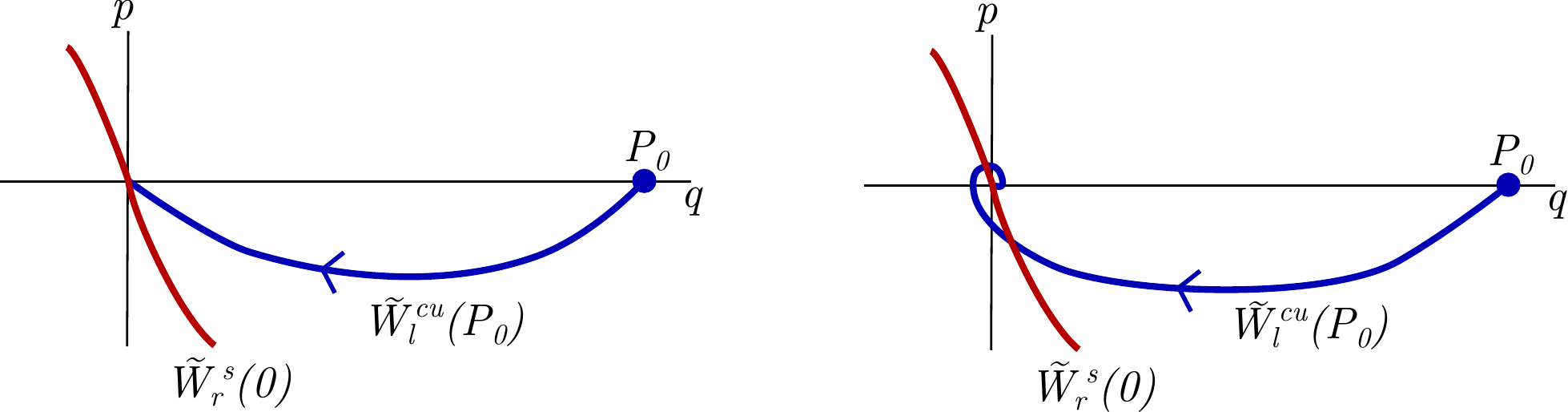}
\caption{Leading order phase portrait of real subspace in \eqref{e:red1} for $\tl c> 4$ (left) and $\tl c\lesssim 4$ (right).}
\label{f:3}
\end{figure}


We now claim that this intersection is transversely unfolded in $\tl\gamma$.  That is, after appending the equation $\fr{d}{d\zeta} \tl \gamma =0$ to \eqref{e:red1}, the corresponding extended manifolds 
$$
\tl W_{\mathrm{r},\mathrm{ext}}^\rs(I\times0) = \{ (\tl \gamma,(p,q)): \tl \gamma\in I, (p,q)\in \tl W^\rs_\mathrm{r}(0)\},\quad \tl W_{\mathrm{l},\mathrm{ext}}^\rcu(I\times \mc{P}) = \{ (\tl \gamma,(p,q)): \tl \gamma\in I, (p,q)\in \tl W^\rcu_\mathrm{l}(\mc{P})\},
$$
with $I\subset \R$ a neighborhood of $0$, intersect transversely in the extended phase space $\R\times\C^2$.  Momentarily assuming this claim, the leading order intersection $\tl W_\mathrm{l}^\rcu(\mc{P})\cap \tl W_\mathrm{r}^\rs(0)$ in \eqref{e:red1} must then persist for $\mc{O}(\epsilon)$-perturbations, implying that intersections persist in the system \eqref{e:red}. The foliation results of Proposition \ref{p:fol} then lift this intersection to the full dynamics, implying that $W^\rs_\mathrm{r}(0)$ and $W^\rcu_\mathrm{l}(\mc{P})$ intersect, from which we conclude the existence of a planar trigger front solution $W^*(x,\xi)$ of \eqref{e:fft}, where once again $W^*(x,0)$ is set to be at the intersection point.  To then obtain the $2\pi/\omega$-time-periodic solutions $u^*_\epsilon(\xi,t)$, we simply set
$$
u^*_\epsilon( \xi,  t) = W^*_\epsilon(\xi + c t ,\xi), \quad \omega = ck.
$$

Hence it only remains to prove the extended transversality claim.  We prove this by using a Melnikov type calculation to show that as $\gamma$ is varied from 0 the invariant manifolds $\tl W^\rcu_\mathrm{l}(\mc{P})$ and $\tl W^\rs_\mathrm{r}(0)$ split with non-zero speed.  

 First, we decompose \eqref{e:red1} into real and imaginary variables $q = q_r + \ri q_i, p = p_r + \ri p_i$,
\begin{align}
\dot{q_r}&= p_r,\notag\\
\dot{p_r}&=-\fr{1}{4}(\mu(\xi)q_r + cp_r -3 q_r(q_r^2 + q_i^2)) - 2\gamma p_i,\notag\\
\dot{q_i}&= p_i\notag\\
\dot{p_i}&= -\fr{1}{4}(\mu(\xi)q_i + cp_i - 3q_i(q_r^2+q_i^2)) - 2\gamma p_r,
\end{align}
For short-hand we denote this system as 
$
U_\zeta = F(\xi,U;\tl c,\tl \gamma)$ with $U\in \R^4.$
First note that the heteroclinic front, $U_*(\zeta) = (q_{*}(\zeta),p_{*}(\zeta),0,0)^T$ found above in \eqref{e:red1} for $\gamma = 0$, lies in the real subspace $\{q_i = p_i = 0\}$.  We wish to study how the invariant manifolds vary in $\gamma$ near $U^*(0)$ so we study the variational equation about $U_*$, 
\begin{align}
V_\zeta &= A(\zeta) V + G(\zeta,V;\tl c,\tl\gamma),\\
A(\zeta) &= D_UF(\zeta,U_*(\zeta);\tl c,0),\quad G(\zeta,V;\tl c,\tl\gamma) = F(\zeta,U_*(\zeta) + V;\tl c, \tl \gamma) - F(\zeta,U_*(\zeta);\tl c,0) - D_UF(\zeta,U_*(\zeta);\tl c,0)V\notag.
\end{align}
It then readily follows that the linear variational equation $V_\zeta= A(\zeta)V$ possesses exponential dichotomies $ \Phi_\mathrm{r}^{\rs/\ru}(\zeta,s)$ for $\zeta,s>0$ and $\Phi_\mathrm{l}^{\rcu/\rss}(\zeta,s)$ for $\zeta,s<0$, with decay properties determined by the linearizations about the asymptotic equilibria; see \cite{sandstede2004defects,sandstede2001structure} for a precise definition.  The corresponding subspaces satisfy
\begin{align}
\tl E_\mathrm{r}^{\rs/\ru}(\zeta)&:=\mathrm{Rg}\Phi_\mathrm{r}^{\rs/\ru}(\zeta,\zeta),\quad
 \tl E_\mathrm{r}^{\rs/\ru}(\zeta) = T_{U^*(\zeta)}\tl W^{\rs/\ru}_\mathrm{r}(0),\quad \zeta\geq0,\notag\\
\tl E_\mathrm{l}^{\rss/\rcu}(\zeta)&:=\mathrm{Rg}\Phi_\mathrm{l}^{\rss/\rcu}(\zeta,\zeta)\quad 
\tl E_\mathrm{r}^{\rss/\rcu}(\zeta) = T_{U^*(\zeta)}\tl W^{\rss/\rcu}_\mathrm{l}(\mc{P}_0),\quad \zeta\leq0.\notag
\end{align}
Furthermore, in a neighborhood of $U^*(0)$, the invariant manifolds can be described by the sets 
\begin{align}
\tl W^\rcu_\mathrm{l}(\mc{P}_0)&=\{U^*(0) + v_\mathrm{l} + \tl h_\mathrm{l}^\rcu(v_\mathrm{l},\tl\gamma)\,|\, 
\tl h_\mathrm{l}^\rcu:\tl E^\rcu_\mathrm{l}(0)\times \R\rightarrow \tl E^\rss_\mathrm{l}(0)\},\notag\\
\tl W^\rs_\mathrm{r}(0)&:= \{ U^*(0) + v_\mathrm{r} + \tl h_\mathrm{r}^\rs(v_\mathrm{r},\tl\gamma)\,|\,
\tl h_\mathrm{r}^\rs: \tl E^\rs_\mathrm{r}(0)\times \R\rightarrow \tl E^\ru_\mathrm{r}(0)\}.\notag
\end{align}
Then since there exists a one parameter family of intersections $\{R(\theta) U^*(0)\}$, where $R(\theta)$ is the real matrix defined by the complex rotation $(q,p)\mapsto \re^{\ri\theta}(q,p)$ in the basis $\{q_r,p_r,q_i,p_i\}$, we have $\mathrm{dim}_{\R^4} \tl E_\mathrm{r}^\rs(0)\cap \tl E_\mathrm{l}^\rcu(0) = 1,$ and thus $\mathrm{codim}_{\R^4} \,\tl E_\mathrm{r}^\rs(0)+ \tl E_\mathrm{l}^\rcu(0) = 1.$  Then using the fact that $\p_\theta R(0) U_*(0)  = (0,0,-q_*(0),-p_*(0))^T$, an explicit calculation gives that the vector $\tl\psi_0:= (0,0,-p_*(0),q_*(0))^T$ satisfies
\begin{equation}
 \mathrm{span}\{\tl\psi_0\}\, = \lp[\tl E_\mathrm{r}^\rs(0)+ \tl E_\mathrm{l}^\rcu(0) \rp]^\perp.\notag
 \end{equation}
We then define a splitting distance for these invariant manifolds in a neighborhood of $U^*(0)$, 
$$
S(\tl\gamma,v_\mathrm{l},v_\mathrm{r}) = \la  \tl\psi_0, \tl h_\mathrm{l}^\rcu(v_\mathrm{l};\tl\gamma) - \tl h_\mathrm{r}^\rs(v_\mathrm{r},\tl\gamma) \ra_{\R^4},
$$
for $v_\rrl$ in some small neighborhood of zero, where here $\la\cdot,\cdot\ra_{\R^4}$ denotes the regular inner product in $\R^4$. Using the variation of constants formula for the invariant manifolds we find
\begin{align}
\fr{\p}{\p\tl\gamma} S(0,0,0) &= \la  \tl\psi_0, \tl h_\mathrm{l}^\rcu(0;0) - \tl h_\mathrm{r}^\rs(0;0) \ra_{\R^4}\notag\\
&= \la \tl\psi_0 ,\int_{-\infty}^0 \Phi_\mathrm{l}^\rss(0,\zeta)\p_\gamma G d\zeta -\int_\infty^0\Phi_\mathrm{r}^\ru(0,\zeta)\p_\gamma G d\zeta    \ra_{\R^4},
\end{align}
where 
$$
\p_\gamma G = \fr{\p}{\p\tl\gamma} G(0,\tl c,0) = (0,-2p^*(0),0,2p^*(0))^T.
$$
We then decompose this inner product and analyze each term separately.  We first find
$$
\int_\infty^0\la \tl\psi_0,\Phi_\mathrm{r}^\ru(0,\zeta)\p_\gamma G d\zeta \ra_{\R^4}< 0,
$$
by approximating $\Phi_\mathrm{r}^\ru(0,\zeta)$ by the linear flow of the asymptotic linearization about $U = 0$.  We next find 
$$
\int_{-\infty}^0 \la \tl\psi_0,\Phi_\mathrm{l}^\rss(0,\zeta)\p_\gamma G \ra_{\R^4} d\zeta >0,
$$
using a phase-plane analysis of the linearized flow about $U^*(\zeta)$ in $\{q_i,p_i\}$ and the invariance of the real and imaginary subspaces under the flow for $\gamma = 0$.  We therefore have
$$
\fr{\p}{\p\tl\gamma} S(0,0,0)>0.
$$

From this we conclude that in a neighborhood of $U^*(0)$ the invariant manifolds $\tl W^\rcu_\mathrm{r}(\mc{P}_0)$ and $\tl W^\rs_\mathrm{r}(0)$ split for $\gamma\neq0$ and thus are transversely unfolded in $\tl\gamma$. This completes the proof of the theorem.
\end{Proof}

Using the transversality obtained in the above proof we can then prove genericity of the front $u_\epsilon^*$.

\begin{Proposition}[Genericity]\label{p:on-st}
The trigger front $u^*_\epsilon$ is non-degenerate in the sense of Definition \ref{d:p}.  That is the linearization $\mc{L}$ about such a front has an algebraically simple Floquet exponent at $\lambda = 0$ when considered on the exponentially weighted space $L^2_\eta(\R\times \mb{T})$ for $\eta>0$, small.
\end{Proposition}
\begin{Proof}
This result can be obtained by viewing the front as a heteroclinic orbit in a $\xi$-spatial dynamics formulation. Given the extended transversality found in Theorem \ref{t:on-ex} above, and the spectral stability of the asymptotic states $u\equiv0$ and $u_\mathrm{p}$ (see \cite{mielke1997instability}), one can use perturbation arguments in $\epsilon$ (see \cite{sandstede1999essential} or \cite{gsu} for example) to observe that the invariant manifolds of the asymptotic states $W_\mathrm{l}^\rcu(u_\mathrm{p})$ and $W_\mathrm{r}^\rs(0)$ in this spatial dynamics formulation are transversely unfolded near $u_\epsilon^*$ in the parameter $\omega$.  Then the approach of \cite{sandstede2004defects} can be used, viewing $u_\epsilon^*$ as a ``transverse transmission" defect, to conclude algebraic simplicity of the zero eigenvalue.  
 \end{Proof}
 
 Combining the results of Theorem \ref{t:on-ex} and Proposition \ref{p:on-st} we obtain the existence of a planar trigger front in the Swift-Hohenberg equation at onset satisfying the necessary hypotheses for Theorem \ref{t:2}.  Hence we can conclude the existence of obliquely striped trigger fronts in \eqref{e:par} at onset.

\section{Applications and discussion}\label{s:dis}

We discuss our results with possible extensions, and mention several applications. 

\paragraph{Summary of results.}
We established existence of coherent structures that create striped patterns at a small oblique angle to a quenching line, in the prototypical model of the Swift-Hohenberg equation. Technically, the result is a singular perturbation result in the presence of continuous spectrum, perturbing from coherent pattern-forming fronts that create stripes parallel to the quenching line. In the second part of this paper, we establish the existence of these primary pattern-forming fronts.

\paragraph{Extending in parameter space.}
Ignoring the dependence on the parameter $\mu$, our results establish existence for speeds $0<c<c_\mathrm{lin}$ and  $0<k_y<k_y(c)$. In a different direction, one could strive to establish existence for $k_y\neq 0$, $0<c<c_\mathrm{max}(k_y)$. As in the case of stripe formation parallel to the quenching line, one expects the maximal speed to be determined by a linear spreading speed, $c_\mathrm{max}(k_y)=c_\mathrm{lin}(k_y)$, that is, the speeds at which disturbances spread by means of pulled invasion fronts \cite{van2003front} when modulated in the transverse direction. It turns out that for isotropic systems, such transversely modulated invasion speeds are always slower than non-modulated invasion, $c_\mathrm{lin}(k_y)<c_\mathrm{lin}(0)$ for all $k_y\neq 0$;  \cite[\S 7]{holzer2014criteria}. We therefore expect that, proving existence of transversely modulated invasion fronts together with a gluing result as in  \cite{goh2014triggered} would give existence of oblique fronts up to $c_\mathrm{max}(k_y)=c_\mathrm{lin}(k_y)$, for a rather general class of pattern-forming equations. The fact that oblique fronts exist up to a maximal speed slower than the maximal speed of parallel fronts then explains the fact that, across many experiments from reaction-diffusion settings to phase separation processes, one observes stripes parallel to the quenching line at large speeds. 

For small speeds, at finite $\mu$, wavenumber selection turns out to be quite intricate \cite{goh2016universal}, and extensions to nonzero $k_y$ appear to be difficult. Beyond small angles, $k_y\sim 0$, one can still attempt to construct stripe-forming fronts for small $\mu$, using the approach from Section \ref{s:on} and more generally spatial dynamics and normal forms as in \cite{wugb}. We expect more intricate behavior when  $k_y$ approaches the zigzag-critical wavenumber, at which point we expect $k_x\sim 0$. It would clearly be interesting to establish existence and non-existence of solutions globally in the space of wavenumbers $k_x,k_y$ and speeds $c$.

\paragraph{Stability, modulation, and selection.}
Beyond existence of solutions, the next natural question would be concerned with their stability. One would expect linear stability of the waves, based on the approximation by a monotone solution in the triggered Ginzburg-Landau amplitude equation. Having established spectral stability in such a perturbative fashion, one expects nonlinear diffusive stability from results for coherent structures as in \cite{gsu}. A difficulty occurs, however, when considering perturbations that are not co-periodic in the $y$-direction. Already linear stability of the selected stripe solution with respect to such perturbations is not evident. As demonstrated in \cite{weinburd}, relying on 
\cite{goh2016universal}, parallel stripes selected by slowly moving triggers are unstable against transverse perturbations due to a zigzag instability. The instability may however be propagating at a slower speed than the quenching, such that unstable patterns can be observed in a large region in the wake of the quenching line. Similar formation of unstable, or merely metastable patterns  in the wake of the quenching line has been observed in phase separation processes; see \cite{monteiro2016} and references therein. It is worth noticing that our perturbation analysis is insensitive to these instabilities, as the main non-degeneracy assumption only relies on co-periodic stability of the periodic patterns. Other instabilities are non-resonant with the mode of propagation and therefore do not change the Fredholm index; see \cite[\S 4]{gms} for the notion of non-resonance in this context. 

On the other hand, the fact that stripe formation occurs for a family of wavenumbers $k_y\sim 0$ raises the question of possible interaction of wavenumbers, modulating an initial condition slowly in the transverse direction along stripe-forming fronts with wavenumber $k_y(\epsilon y)$. Focusing on the dynamics at the interfacial line and neglecting the farfield patterns, one then observes temporal oscillations that are slowly modulated in $y$ and expects dynamics for the wavenumber $k_y$ given by Burgers' equation \cite{dsss}. It would be interesting to understand if and how these interfacial dynamics interact with the far-field striped patterns. Beyond $k_y\sim 0$, one observes more complicated dynamics, such as interaction between stripes formed parallel to the interface in $y>0$, say, and stripes perpendicular to the interface in $y<0$, say, creating permanent grain boundaries between different stripe orientation in their wake \cite{lloyd2016continuation,wugb}.

\begin{figure}[h!]\centering
\includegraphics[width=.8\textwidth]{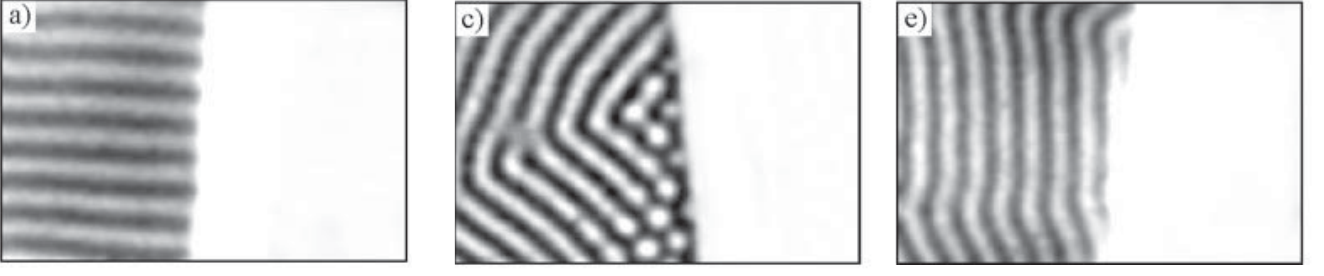}\\
\includegraphics[clip,trim={0 1.75in 0.1in 1.75in},width=0.33\textwidth]{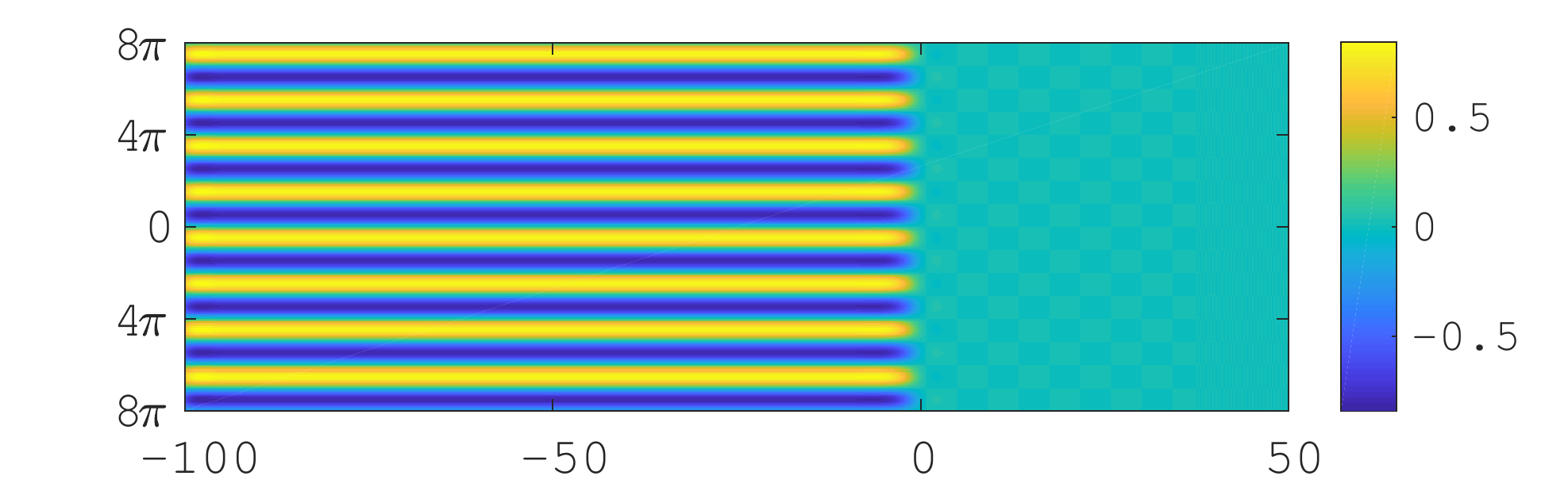}\hspace{-0.4in}
\includegraphics[clip,trim={0 1.75in 0.1in 1.75in},width=0.33\textwidth]{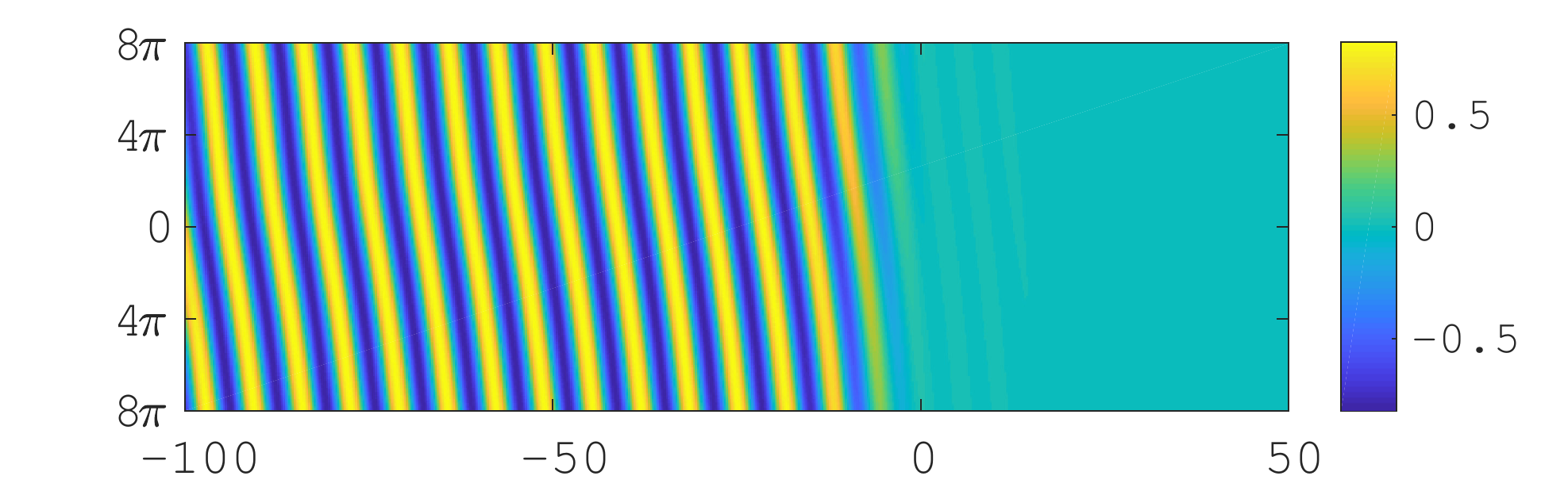}\hspace{-0.4in}
\includegraphics[clip,trim={0 1.75in 0.1in 1.75in},width=0.33\textwidth]{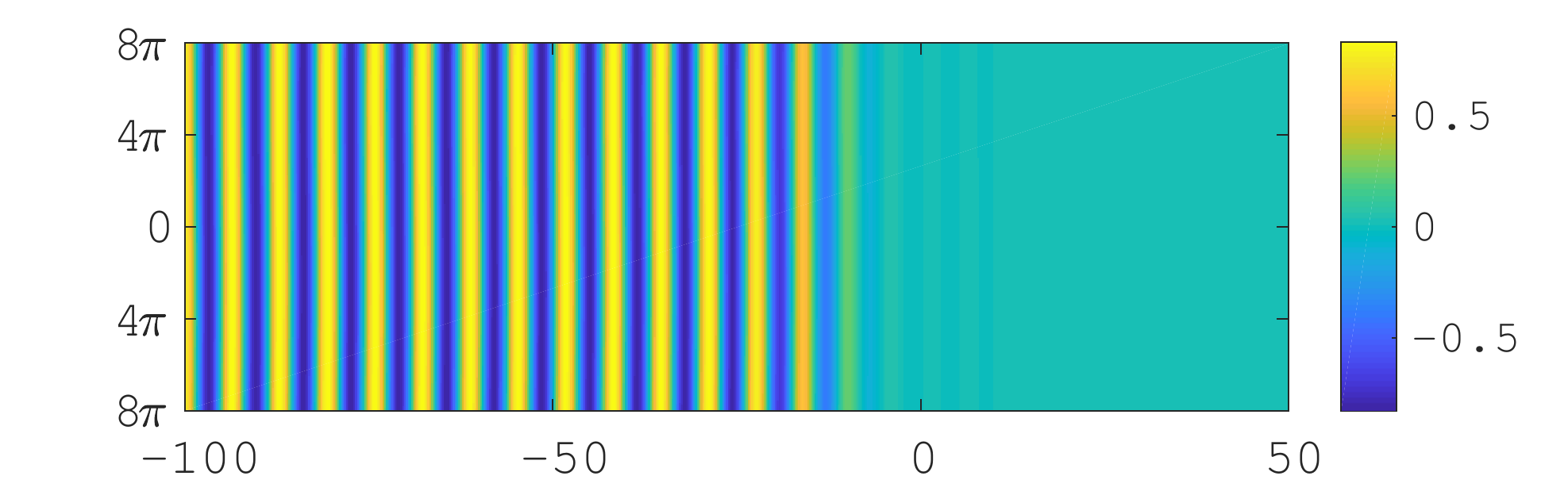}
\caption{Top row: Time snap shots of striped patterns of different orientations in a triggered light sensitive reaction-diffusion system for increasing excitation speeds (from left to right); figure adapted with permission from  \cite{miguez2006effect}. Copyrighted by the American Physical Society. Bottom row:  Simulations of \eqref{e:sh0} with $\mu_0 = 0.5$ and random initial data for increasing speeds $c = 0.5, 2, 2.5$ at $t = 500.$ Simulated on the computational domain $(\xi,y)\in [-80\pi,80\pi]\times[-8\pi,8\pi]$ with periodic boundary conditions.  We used an exponential time differencing scheme with $dt = 0.1$, and a first order spectral method in space with $2^{11}$ and $2^8$ modes in $\xi$ and $y$ respectively.  The unstable domain where $\mu(\xi)>0$ was cut-off near the left boundary (not pictured here) to satisfy the boundary conditions.}
\label{f:2}
\end{figure}

\paragraph{Experiments.}
We briefly report on experiments exhibiting the parallel and oblique stripe formation. Considering the Swift-Hohenberg equation as a prototypical model for Turing instabilities leading to striped phases, the most relevant recent experiments have been performed using a light-sensing reaction-diffusion system, striving to illustrate the effect of domain growth on the formation of Turing patterns \cite{miguez2006effect}. More specifically, the authors studied a CDIMA reaction in a gel in which high intensity light suppresses a spatial Turing instability.  They found that by moving an opaque mask, which blocked the light, across the surface of the gel at different speeds, different patterns were selected. Stripes parallel to the mask boundary were observed for relatively fast mask speeds, oblique stripes were observed for moderate speeds, and stripes perpendicular to the interface were observed for slow speeds, in agreement with our predictions that oblique stripe formation is compatible only with speeds less than the maximal speed of parallel stripe formation as discussed above; see Figure \ref{f:2}. This experimental set-up can also be viewed as a caricature of pattern formation in growing domains, where boundary conditions and growth rate can select patterns in the bulk \cite{goh2016universal, morrissey2015characterizing}. 

We remark that similar qualitative behavior can be observed in our Swift-Hohenberg model \eqref{e:sh0} as in the aforementioned reaction-diffusion model; compare Fig. \ref{f:2}(a) and (b).We notice that the phenomena are in agreement with our discussion, here. As we explained, we expect stripe formation with wavenumber $k_y$ to be compatible with the quenching process when $c<c_\mathrm{lin}(k_y)$. Since $c_\mathrm{lin}(k_y)$ is quite universally decreasing in $k_y$ \cite{holzer2014criteria}, one expects that for large speeds only small wavenumbers $k_y\sim 0$ nucleate at the quenching interface. 

Another set of examples arise in phase-separating systems which are externally triggered into a bistable regime from a monostable regime in a spatially-progressive manner; see for example \cite{foard2012survey,thomas2013probability,furukawa,kurita}.   Depending on the speed of excitation and characteristics of the system, different orientations of stripes relative to the quenching direction can be observed.  The recent work \cite{monteiro2016} has performed a theoretical classification of such phenomenon in the prototypical Allen-Cahn and Cahn-Hilliard equations posed on the plane.  Of particular relevance to our work, the authors found that obliquely oriented stripes do not exist in the quenched Allen-Cahn equation for both zero and small non-zero quenching rates. Furthermore, they suggested that oblique stripes accompany planar stripes, or ``vertical," stripes in the Cahn-Hilliard equation for non-zero quenching rates, but left the question of rigorous existence unanswered. Our present work, albeit in the Swift-Hohenberg context, suggests such patterns do in fact exist. Together with the construction of stripe-formation parallel to the interface in a quenched Cahn-Hilliard equation based on matched asymptotics \cite{krekhov}, one would hope to be able to adapt our strategy to establish the existence of such fronts forming oblique stripes. Again, the prediction of parallel only stripes at larger speeds that we predict to be quite universally valid is in agreement with the observations, here \cite{foard2012survey,kurita}.

\bibliography{sh-oblique}

\begin{thebibliography}{10}

\bibitem{archer2012solidification}
A.~J. Archer, M.~J. Robbins, U.~Thiele, and E.~Knobloch.
\newblock Solidification fronts in supercooled liquids: How rapid fronts can
  lead to disordered glassy solids.
\newblock {\em Physical Review E}, 86(3):031603, 2012.

\bibitem{crampin1999reaction}
E.~J. Crampin, E.~A. Gaffney, and P.~K. Maini.
\newblock Reaction and diffusion on growing domains: Scenarios for robust
  pattern formation.
\newblock {\em Bulletin of Mathematical Biology}, 61(6):1093--1120, Nov 1999.

\bibitem{doelman2003propagation}
A.~Doelman, B.~Sandstede, A.~Scheel, and G.~Schneider.
\newblock Propagation of hexagonal patterns near onset.
\newblock {\em European Journal of Applied Mathematics}, 14(01):85--110, 2003.

\bibitem{dsss}
A.~Doelman, B.~Sandstede, A.~Scheel, and G.~Schneider.
\newblock The dynamics of modulated wave trains.
\newblock {\em Mem. Amer. Math. Soc.}, 199(934):viii+105, 2009.

\bibitem{eckmann1991propagating}
J.-P. Eckmann and C.~Wayne.
\newblock Propagating fronts and the center manifold theorem.
\newblock {\em Communications in Mathematical Physics}, 136(2):285--307, 1991.

\bibitem{fenichel1977asymptotic}
N.~Fenichel.
\newblock Asymptotic stability with rate conditions. 2.
\newblock {\em Indiana University Mathematics Journal}, 26(1):81--93, 1977.

\bibitem{fenichel1979geometric}
N.~Fenichel.
\newblock Geometric singular perturbation theory for ordinary differential
  equations.
\newblock {\em Journal of Differential Equations}, 31(1):53--98, 1979.

\bibitem{fiedlerscheel}
B.~Fiedler and A.~Scheel.
\newblock Spatio-temporal dynamics of reaction-diffusion patterns.
\newblock In {\em Trends in nonlinear analysis}, pages 23--152. Springer,
  Berlin, 2003.

\bibitem{foard2012survey}
E.~Foard and A.~Wagner.
\newblock Survey of morphologies formed in the wake of an enslaved
  phase-separation front in two dimensions.
\newblock {\em Physical Review E}, 85(1):011501, 2012.

\bibitem{furukawa}
H.~Furukawa.
\newblock Phase separation by directional quenching and morphological
  transition.
\newblock {\em Physica A: Statistical Mechanics and its Applications},
  180(1):128 -- 155, 1992.

\bibitem{gsu}
T.~Gallay, G.~Schneider, and H.~Uecker.
\newblock Stable transport of information near essentially unstable localized
  structures.
\newblock {\em Discrete Contin. Dyn. Syst. Ser. B}, 4(2):349--390, 2004.

\bibitem{PhysRevE.92.042602}
K.~Glasner.
\newblock Hexagonal phase ordering in strongly segregated copolymer films.
\newblock {\em Phys. Rev. E}, 92:042602, Oct 2015.

\bibitem{goh2016universal}
R.~Goh, R.~Beekie, D.~Matthias, J.~Nunley, and A.~Scheel.
\newblock Universal wave-number selection laws in apical growth.
\newblock {\em Phys. Rev. E}, 94:022219, Aug 2016.

\bibitem{goh2014triggered}
R.~Goh and A.~Scheel.
\newblock Triggered fronts in the complex {G}inzburg {L}andau equation.
\newblock {\em Journal of Nonlinear Science}, 24(1):117--144, 2014.

\bibitem{gms}
R.~N. Goh, S.~Mesuro, and A.~Scheel.
\newblock Spatial wavenumber selection in recurrent precipitation.
\newblock {\em SIAM J. Appl. Dyn. Syst.}, 10(1):360--402, 2011.

\bibitem{huaruagucs1999bifurcating}
M.~H{\u{a}}r{\u{a}}gu{\c{s}}-Courcelle and G.~Schneider.
\newblock Bifurcating fronts for the {T}aylor--{C}ouette problem in infinite
  cylinders.
\newblock {\em Zeitschrift f{\"u}r angewandte Mathematik und Physik},
  50(1):120--151, 1999.

\bibitem{holzer2014criteria}
M.~Holzer and A.~Scheel.
\newblock Criteria for pointwise growth and their role in invasion processes.
\newblock {\em Journal of Nonlinear Science}, 24(4):661--709, 2014.

\bibitem{kato2013perturbation}
T.~Kato.
\newblock {\em Perturbation theory for linear operators}, volume 132.
\newblock Springer Science \& Business Media, 1966.

\bibitem{kopf2014emergence}
M.~H. K{\"o}pf and U.~Thiele.
\newblock Emergence of the bifurcation structure of a {L}angmuir--{B}lodgett
  transfer model.
\newblock {\em Nonlinearity}, 27(11):2711, 2014.

\bibitem{krekhov}
A.~Krekhov.
\newblock Formation of regular structures in the process of phase separation.
\newblock {\em Phys. Rev. E}, 79:035302, Mar 2009.

\bibitem{kurita}
R.~Kurita.
\newblock Control of pattern formation during phase separation initiated by a
  propagated trigger.
\newblock {\em Scientific Reports}, 7(1):6912, 2017.

\bibitem{lloyd2016continuation}
D.~J.~B. Lloyd and A.~Scheel.
\newblock Continuation and bifurcation of grain boundaries in the
  {S}wift-{H}ohenberg equation.
\newblock {\em SIAM J. Appl. Dyn. Syst.}, 16(1):252--293, 2017.

\bibitem{mielke1997instability}
A.~Mielke.
\newblock Instability and stability of rolls in the {S}wift-{H}ohenberg
  equation.
\newblock {\em Comm. Math. Phys.}, 189(3):829--853, 1997.

\bibitem{miguez2006effect}
D.~G. M{\'\i}guez, M.~Dolnik, A.~P. Mu{\~n}uzuri, and L.~Kramer.
\newblock Effect of axial growth on {T}uring pattern formation.
\newblock {\em Physical Review Letters}, 96(4):048304, 2006.

\bibitem{monteiro2016}
R.~Monteiro and A.~Scheel.
\newblock Phase separation patterns from directional quenching.
\newblock {\em Journal of Nonlinear Science}, Feb 2017.

\bibitem{morrissey2015characterizing}
D.~Morrissey and A.~Scheel.
\newblock Characterizing the effect of boundary conditions on striped phases.
\newblock {\em SIAM Journal on Applied Dynamical Systems}, 14(3):1387--1417,
  2015.

\bibitem{pennybacker2013phyllotaxis}
M.~Pennybacker and A.~C. Newell.
\newblock Phyllotaxis, pushed pattern-forming fronts, and optimal packing.
\newblock {\em Physical Review Letters}, 110(24):248104, 2013.

\bibitem{rademacher2007saddle}
J.~D. Rademacher and A.~Scheel.
\newblock The saddle-node of nearly homogeneous wave trains in
  reaction--diffusion systems.
\newblock {\em Journal of Dynamics and Differential Equations}, 19(2):479--496,
  2007.

\bibitem{risler}
E.~Risler.
\newblock Travelling waves and dispersion relation in the spatial unfolding of
  a periodic orbit.
\newblock {\em C. R. Math. Acad. Sci. Paris}, 334(9):833--838, 2002.

\bibitem{robbins2012modeling}
M.~J. Robbins, A.~J. Archer, U.~Thiele, and E.~Knobloch.
\newblock Modeling the structure of liquids and crystals using one-and
  two-component modified phase-field crystal models.
\newblock {\em Physical Review E}, 85(6):061408, 2012.

\bibitem{sandstede1999essential}
B.~Sandstede and A.~Scheel.
\newblock Essential instability of pulses and bifurcations to modulated
  travelling waves.
\newblock {\em Proceedings of the Royal Society of Edinburgh-A-Mathematics},
  129(6):1263--1290, 1999.

\bibitem{sandstede2001structure}
B.~Sandstede and A.~Scheel.
\newblock On the structure of spectra of modulated travelling waves.
\newblock {\em Mathematische Nachrichten}, 232(1):39--93, 2001.

\bibitem{sandstede2004defects}
B.~Sandstede and A.~Scheel.
\newblock Defects in oscillatory media: toward a classification.
\newblock {\em SIAM Journal on Applied Dynamical Systems}, 3(1):1--68, 2004.

\bibitem{sandstede2008relative}
B.~Sandstede and A.~Scheel.
\newblock Relative {M}orse indices, fredholm indices, and group velocities.
\newblock {\em Discrete and Continuous Dynamical Systems A}, pages 139--158,
  2008.

\bibitem{weinburd}
A.~Scheel and J.~Weinburd.
\newblock Wavenumber selection in a quenched {S}wift-{H}ohenberg equation.
\newblock {\em {P}reprint}, 2017.

\bibitem{wugb}
A.~Scheel and Q.~Wu.
\newblock Small-amplitude grain boundaries of arbitrary angle in the
  {S}wift-{H}ohenberg equation.
\newblock {\em ZAMM Z. Angew. Math. Mech.}, 94(3):203--232, 2014.

\bibitem{thomas2013probability}
S.~Thomas, I.~Lagzi, F.~Moln{\'a}r~Jr, and Z.~R{\'a}cz.
\newblock Probability of the emergence of helical precipitation patterns in the
  wake of reaction-diffusion fronts.
\newblock {\em Physical Review Letters}, 110(7):078303, 2013.

\bibitem{van2003front}
W.~Van~Saarloos.
\newblock Front propagation into unstable states.
\newblock {\em Physics Reports}, 386(2):29--222, 2003.

\end{thebibliography}
\bibliographystyle{abbrv}

\Addresses

\end{document}